\RequirePackage{fix-cm}
\documentclass[twocolumn]{svjour3}          
\smartqed  

\usepackage{graphicx}
\usepackage{amsmath}
\usepackage{subfigure}
\usepackage{cite}
\usepackage{units}
\usepackage{pifont}
\usepackage{enumerate,setspace,mathtools,amssymb,array,todonotes,multirow}

\usepackage[ruled,vlined]{algorithm2e}
\DeclareMathOperator*{\argminC}{\arg\min}

\begin{document}

\title{NeuroPath2Path: Classification and elastic morphing between neuronal arbors using path-wise similarity
}

\author{Tamal Batabyal         \and
        Barry Condron  \and
	Scott T. Acton  
}


\institute{Tamal Batabyal \at
              Department of Electrical \& Computer Engineering \\
	     University of Virginia \\	
              Tel.: +1 4344665486\\              
              \email{tb2ea@virginia.edu}           
           \and
           Barry Condron \at
              Department of Biology\\
	      University of Virginia 
	   \and
	  Scott T. Acton \at
		Department of Electrical \& Computer Engineering\\
	      Department of Biomedical Engineering\\
            University of Virginia
}

\date{Received: date / Accepted: date}

\maketitle

\begin{abstract}
The shape and connectivity of a neuron determine its function. Modern imaging methods have proven successful at extracting such information. However, in order to analyze this type of data, neuronal morphology needs to be encoded in a graph-theoretic method. This encoding enables the use of high throughput informatic methods to extract and infer brain function. The application of graph-theoretic methods to neuronal morphological representation comes with certain difficulties. Here we report a novel, effective method to accomplish this task.

The morphology of a neuron, which consists of its overall size, global shape, local branch patterns, and cell-specific biophysical properties, can vary significantly with the cell's identity, location, as well as developmental and physiological state. 
  Various algorithms have been developed to customize shape based statistical and graph related features for quantitative analysis of neuromorphology, followed by the classification of neuron cell types using the features. Unlike the classical feature extraction based methods from imaged or 3D reconstructed neurons, we propose a model based on the rooted path decomposition from the soma to the dendrites of a neuron and extract morphological features on each path. We hypothesize that measuring the distance between two neurons can be realized by minimizing the cost of continuously morphing the set of all rooted paths of one neuron to another. To validate this claim, we first establish the correspondence of paths between two neurons using a modified Munkres algorithm. Next, an elastic deformation framework that employs the square root velocity function is established to perform the continuous morphing, which, in addition, provides an effective visualization tool. We experimentally show the efficacy of NeuroPath2Path,  \textit{NeuroP2P}, over the state of the art.   
\end{abstract}

\section{Introduction}
\label{intro}
Neurons process information by transmitting electrical signals via complex  circuitry.  
The functionality of each neuron depends on a set of intrinsic factors, such as morphology, ionic channel density, gene expression, including the extrinsic ones, such as connectivity to other neurons~\cite{brown2008quantifying, ascoli2008petilla}. In 1899, Cajal~\cite{y1972histologie}, considered the founder of modern neuroscience, put forward his pioneering work on neuroanatomy with detailed, accurate, and meticulous illustrations, and posited that the shape of a neuron determines its functionality.  Experimental results strongly support this idea. Inspired by this fundamental work, the study of neuromorphology primarily aims at analyzing and quantifying the complex shape and physiology of neurons in specific functional regions to identify relationships. 

\begin{figure*}[t]
\vspace{-.2cm}	
	\centering
	\renewcommand{\tabcolsep}{0.05cm}
\begin{tabular}{cc@{\hskip 0.15cm}cc}
	{\includegraphics[width=4.2cm, height=2.5cm]{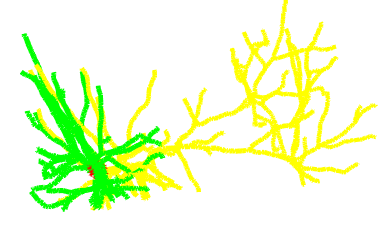}}&
	{\includegraphics[width=4.2cm,height=2.5cm]{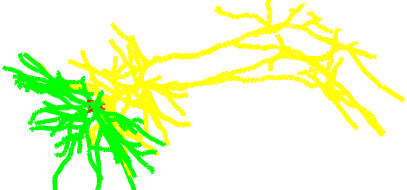}}&
    {\includegraphics[width=4.2cm,height=2.5cm]{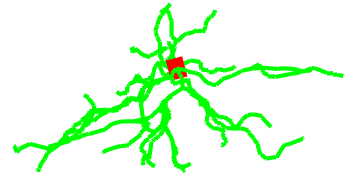}}&
    {\includegraphics[width=4.2cm,height=2.5cm]{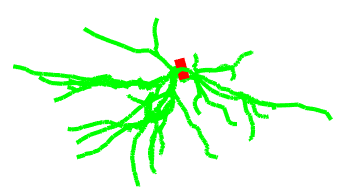}}\\
    \multicolumn{2}{c}{(a)} & \multicolumn{2}{c}{(b)}\\
    {\includegraphics[width=4.2cm,height=2.5cm]{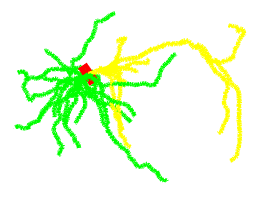}}&
    {\includegraphics[width=4.2cm,height=2.5cm]{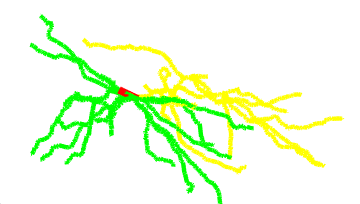}}&
    {\includegraphics[width=4.2cm,height=2.5cm]{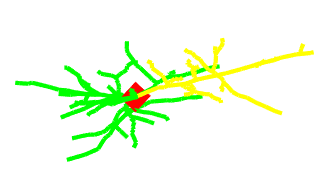}}&
    {\includegraphics[width=4.2cm,height=2.5cm]{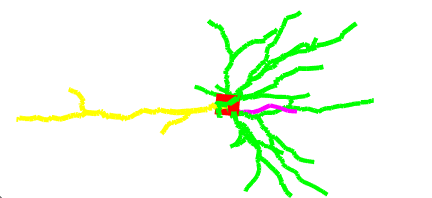}}\\
    \multicolumn{2}{c}{(c)} & \multicolumn{2}{c}{(d)}\\
    {\includegraphics[width=4.2cm,height=2.5cm]{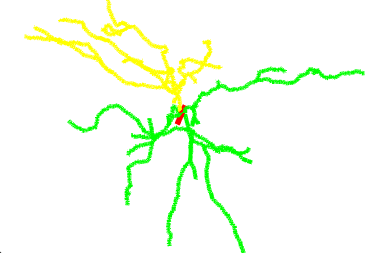}}&
    {\includegraphics[width=4.2cm,height=2.5cm]{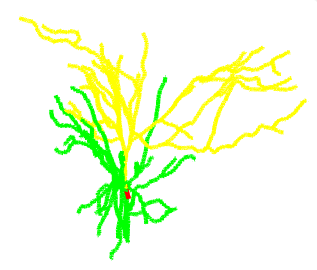}}&
    {\includegraphics[width=4.2cm,height=2.5cm]{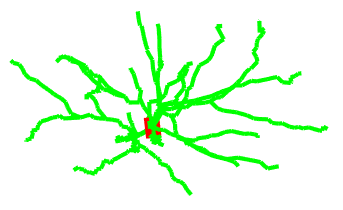}}&
    {\includegraphics[width=4.2cm,height=2.5cm]{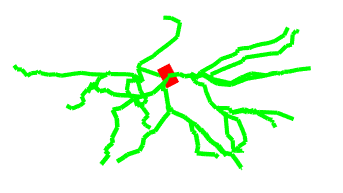}}\\
    \multicolumn{2}{c}{(e)} & \multicolumn{2}{c}{(f)}
\end{tabular}
	\caption{ Pyramidal neurons from (a) primary motor cortex, (b) secondary motor cortex, (c) prefrontal cortex, (d) somatosensory cortex, (e) primary visual cortex, and (f) secondary visual cortex of the mouse. The dendritic branches in yellow are apical dendrites, and in green are basal dendrites. The red square in each cell is the soma, used as the designated root node in our analysis.
	This figure provides a glimpse of region-based arborial differences among pyramidal cells. Cells differ in size and volume, which are scaled for visualization.   }
	\label{fig: Pyrmd}
	\vspace{-.4cm}
\end{figure*} 

Neurons vary significantly in size, shape, and length. A major obstacle towards understanding the brain is the development of efficient ways to encode these shapes. 
The anatomical and geometrical features of neurons of any cell-type, for example, pyramidal cells differ based on the regions in which the cells reside~\cite{bielza2014branching, brown2008quantifying}. 
Fig.~\ref{fig: Pyrmd} shows regional variation in the structure and geometry of dendritic arbors of pyramidal cells~\cite{romand2011morphological}. 
It is observed that the number of branches, length, surface area, and volume of apical dendrites is $4-9$ times larger for hippocampal than for cortical regions, whereas in terms of the same features of basal dendritic arbors, it is approximately $3$ times~\cite{brown2008quantifying}. Another source of variation stems from technical imprecision in measurements obtained while performing 3D reconstruction from image stacks using software tracing tools, such as Neurolucida~\cite{glaser1990neuron}. Noise due to technical imprecision includes wide variations in the number of manually or semi-automatically traced 3D locations (approximately between $60$ to $70,000$), the number of ramified branches and bifurcations by different tracers, and deletion of dendritic spines adversely affect the registration of neurons, and thereby induce error in morphological feature quantification. 
The skeletons of dendritic and axonal branches form a tree topology with a number of bifurcations. The bifurcations at successive stages help in a series of effective and unambiguous signal processing modules, such as active and passive signal propagation, integration, filter, attenuation, oscillation, and backpropagation~\cite{ascoli2008petilla, london2005dendritic}.

\begin{figure*}[t]
\vspace{-.2cm}	
	\centering
	\subfigure[]{\includegraphics[width=5.2cm, height=3.5cm]{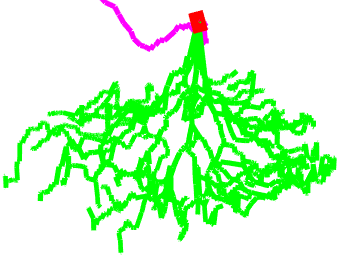}}
	\subfigure[]{\includegraphics[width=5.2cm, height=3.5cm]{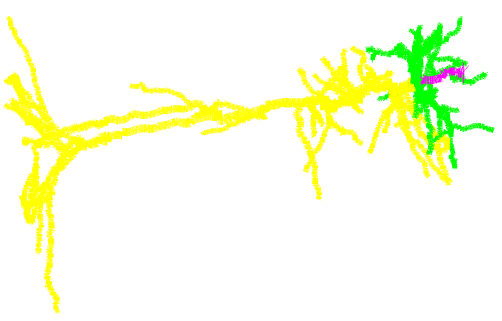}}
	\subfigure[]{\includegraphics[width=5.2cm, height=3.5cm]{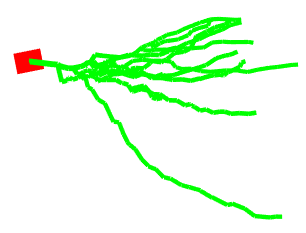}}
	\subfigure[]{\includegraphics[width=5.2cm, height=3.5cm]{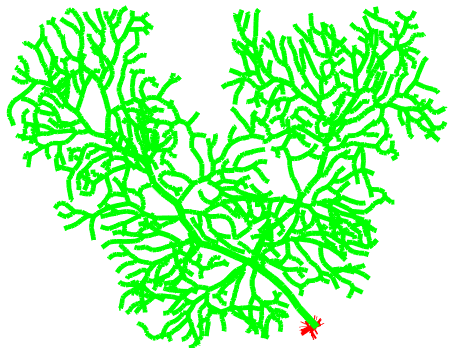}}
	\subfigure[]{\includegraphics[width=5.2cm, height=3.5cm]{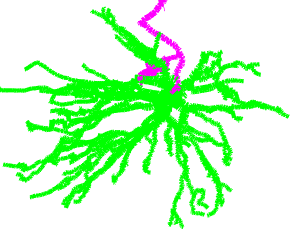}}
	\subfigure[]{\includegraphics[width=5.2cm, height=3.5cm]{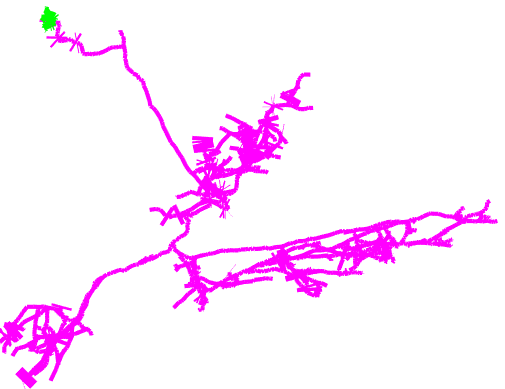}}
	\caption{ (a) A retinal ganglion cell from the inner plexiform layer of a $9$ month-old adult mouse. The 3D reconstructed cell has $3938$ 3D locations, $50$ bifurcations, $106$ branches, $56$ rooted paths, $255.52\mu m$ height, $4499.5\mu m$ diameter, $9061.14\mu m^3$ volume, $18,213.9\mu m^2$ surface area. (b) A 3D reconstructed (traced by Neuromantic~\cite{myatt2012neuromantic}) pyramidal cell of an adult mouse having $24,868$ 3D locations, $95$ bifurcations, $200$ branches, $106$ tips, $814.05\mu m$ height, $16341.8\mu m$ diameter, $12569 \mu m^3$ volume, $24,825.9\mu m^2$ surface area. (c) A hippocampal granule cell (in the dentate gyrus) of a $9$ month-old mouse is traced using Neurolucida. The 3D reconstructed neuron contains $414$ 3D locations with $7$ bifurcations, $15$ branches, $14$ rooted paths, $98.33 \mu m$ height, $443.18\mu m$ diameter, $3107.27\mu m^3$ volume, and $3255.33\mu m^2$ surface area. (d) A Purkinje cell in cerebellar cortex of a $28$ day-old mouse. The reconstruction is performed by Neurolucida, containing $3187$ 3D locations. The neuron has $391$ bifurcations, $783$ branches, $393$ tips, $184.35 \mu m$ height, $4366.24 \mu m$ diameter, $12,794.7\mu m^3$ volume, and $31,574.8\mu m^2$ surface area. (e) A motor cell in the spinal cord of a $10$ day-old mouse, which is reconstructed with $5868$ locations using Neurolucida tracer. The 3D traced neuron has $47$ bifurcations, $103$ branches, $58$ tips, $415.6 \mu m$ height, $784.158\mu m$ diameter, $4006.15\mu m^3$ volume, and $11,931.8 \mu m^2$ surface area. (f) A long-axon projection neuron from the thalamus of a $6$ months old mouse - it is traced by Large Volume Viewer(LVV)~\cite{murphy2014janelia} with $2818$ 3D locations. The traced neuron has $169$ bifurcations, $265$ branches, $2731.69 \mu m$ height, $55,742.44 \mu m^3$ volume and $189979 \mu m^2$ surface area. The color code is the following: yellow=apical dendrites, green = basal dendrites, magenta = axons, red = cell body/soma/root. The quantified statistics on the number of bifurcations and tips or rooted paths that are mentioned above are extracted from dendritic arbors of each cell-type.     }
	\label{fig3}
	\vspace{-.4cm}
\end{figure*} 

From the soma to the dendritic terminals of a neuron, the diameter of the dendritic shaft tapers~\cite{jan2010branching, wen2008cost}. The increased diameter of a dendritic shaft near the soma is tailored to faster signal propagation to the soma compared to the dendritic tuft , which helps generate action potential in the soma. Several research works consider the branches in the proximity of soma are more important compared to the distant dendritic tuft and spines in the analysis of neuromorphology~\cite{kanari2018topological, cervantes2018morphological, lopez2011models}. The length of dendritic branch segments shows similar behavior when propagating away from soma. For instance, the terminal segments are longer than the intermediate branch segments for basal dendrites in cortical pyramidal cells~\cite{bielza2014branching}. These observations support the Bayesian philosophy which is geared towards the analysis of morphogenesis of neurons~\cite{lopez2011models}. Functions such as synaptic boosting~\cite{migliore2002emerging}, coadaptive local spiking~\cite{gasparini2004initiation}, and global spike amplification~\cite{williams2004spatial} suggest the use of other morphometrics to describe the structural aspects on the functions. For example, packing density of ramified branches and bifurcations of neuron potentially trigger intermittently co-adpative spiking.

The tree-type arbors of neurons and the availability of the inventory of digitally-traced 3D reconstructed neurons, Neuromorpho~\cite{ascoli2007neuromorpho}, provided significant momentum in the last decade for the quantitative and qualitative assessment of neuroanatomy via graph-based morphometrics. In Neuromorpho, the sequentially-aligned slices of microscopic images are registered and traced using software~\cite{meijering2010neuron}, such as Neurolucida~\cite{glaser1990neuron} and Neuromantic~\cite{myatt2012neuromantic}, and the reconstructed images can then be processed through software, such as L-measure~\cite{scorcioni2008measure} to extract an extensive list of morphological metrics. On one hand, there are several research works dedicated to analyze the neuromorphology of specific cell types, such as basal dendrites of cortical pyramidal cells~\cite{lopez2011models, bielza2014branching}, GABAergic interneuron cells~\cite{ascoli2008petilla} and others. These works account for region-specific variations in the physiology and anatomy of a neuron cell to establish the effect of certain functions on the structure. On the other hand, research efforts, such as blastneuron~\cite{wan2015blastneuron}, neurosol~\cite{batabyal2017neurosol},and TMD~\cite{kanari2018topological}, attempt to extract model based features, which are catered to the need for automated classification of different neuronal cells. The motivation behind this avenue of research is that it is impossible to identify and categorize one trillion neuronal cells by adopting manual or even semi-automatic methods.

State-of-the-art methods~\cite{wan2015blastneuron,batabyal2018neurobfd} for the classification of neurons can be broadly divided in two categories. Research in the $first$ category, which is supervised in nature, employs different feature extraction algorithms followed by suitable classifiers to obtain classification accuracy in percentage. The validation of the methods are performed by adopting a series of statistical tests. However, the significant variation in the neuron skeletons precludes the selection of the optimal set of morphometrics as features. Adoption of feature transformations, such as principal component analysis (PCA) or kernel transformations, may improve the classification accuracy. Nevertheless, these transformations obscures the identity of discriminating features as the transformed space is formed by linear or nonlinear composition of extracted features. In addition, the classification accuracy of categorization does not quite explain the physiological and structural differences between two neurons.

The $second$ category follows mostly unsupervised approaches and attempts to compute pairwise distances between neurons. Authors in~\cite{sarkar2013shape} used Fourier based shape descriptors to encode the global shape of a neuron, which however ignored the local features of the neuron arbor. Gillette~\cite{gillette2015topological, gillette2015topological1} performed a sequence alignment based algorithm for categorization by decomposing a neuron into a sequence of branches. The approach failed to consider geometric features. Blastneuron~\cite{wan2015blastneuron} adopted a mixed strategy. Using a supervised approach, the method first extracted $21$ global morphological features and $13$ moment invariant features to retrieve a set of targets that closely matches to each test neuron in terms of the anatomical structures. Each target is then RANSAC sampled~\cite{fischler1981random} and aligned optimally to the test neuron, which outputs a distance value. This unsupervised routine decides the output category of the test neuron based on the minimum distance criteria. The method involved initial pruning of branches and resampling of each neuron, which collectively alters the morphology statistics. Moreover, the retrieval accuracy of $233$ projected neurons (PN) of Drosophila drops significantly to $39$\% as the number of potential candidates that are to be compared with the target increases. NeuroSoL~\cite{batabyal2017neurosol} offered a graph-theoretic method which is free from registration and resampling. In spite of its appeal of using graph theory, the matrix alignment routine is NP-hard in nature, thereby producing suboptimal results. The problem of comparing a pair of neuron topologies can also be regarded as a graph kernel based similarity measure problem~\cite{vishwanathan2010graph}. However, the rationale behind conventional graph kernels, such as the random walk kernel may be inconsistent with the morphological understanding of a neuron.

Instead of modeling a neuron as a generic graph, the neuron can be modeled as a specialized graph that contains a collection of rooted paths, where each path starts from the soma, called the root node, and ends up in a dendritic terminal. It is important to note that each path acts like an \textit{atomic neuron}, as it contains the soma and a dendritic end to complete a circuit. Most of the synapses along a path will be nearer the soma than at the end of the path. 
It is convenient to think problems, such as synaptic plasticity as the evolution of a set of synapses over time along all the paths. During this evolution, there are birth, death and rearrangement of paths. Following the same logic, \textit{quantifying the problem of distinguishing two neurons can be equivalently mapped as finding the cost of evolving a set of circuits optimally from one neuron to the other}.  
\begin{figure*}[t]
\vspace{-.2cm}	
	\centering
	\includegraphics[width=15.6cm, height=8cm]{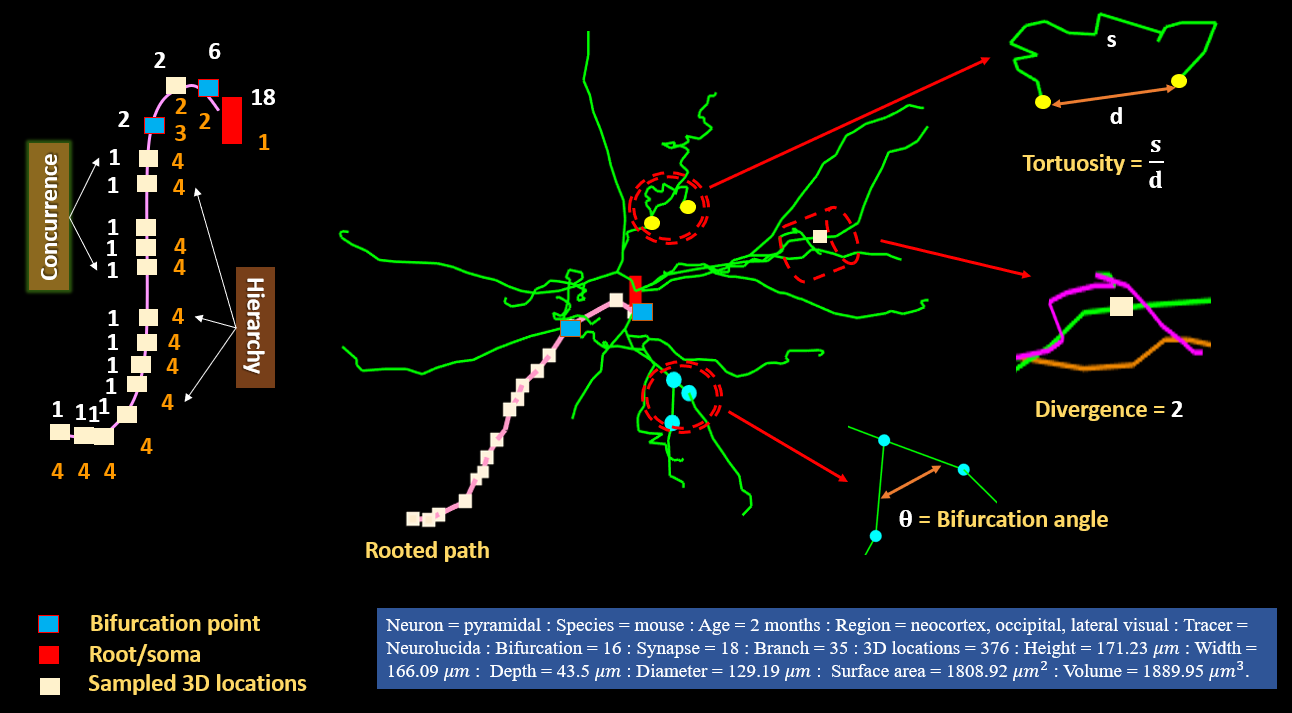}
	\caption{The figure shows a set of morphometrics that are used in our work. The metrics are numerically computed from a neocortical pyramidal cell with its relevant anatomical information provided at the bottom of the figure. A rooted path of the neuron (at the center) is shown (pink) with the 3D locations and bifurcation points, and the corresponding concurrence and hierarchy values are noted on the left hand side of the figure. 
	$18$ paths are originated from the soma, indicating $18$ synaptic terminals. The immediate bifurcation point has a concurrence value of $6$, because, including the current path, there are a total of $6$ paths that end up in $6$ different terminals. Concurrence values of the rest of the 3D locations are computed accordingly. It can be observed that there are four concurrence values that are marked on the chosen path, $\{1,2,6,18\}$. The hierarchy values are obtained by sorting the indices of these four values in reverse order, which in turn describe the depths of locations with respect to the root. On the right hand side, three morphometrics on the selected paths are demonstrated, which are tortuosity, divergence, and bifurcation angle. Quantification of the wrinkle or tangle of a segment (tortuosity) existing between either two consecutive bifurcation points, or a bifurcation point and a following terminal is performed by measuring the curve length $s$, and the Euclidean distance between the start and end locations, $d$. Neocortical pyramidal cells show pronounced wrinkles in their branches. Divergence of a location entails competitive behavior. With a distance scale fixed beforehand, the number of branch segments that are in the immediate neighborhood of a location defines its divergence. Bifurcation angle is another important morphometric, which we measure by using the inner product between two vectors emanating from the bifurcation point. 
	Wide bifurcation angles connote greater exploration of extracellular environment. More branches and smaller bifurcation angles, in general, lead to higher divergence. Neurons with higher divergence tend to have longer path lengths. }
	\label{fig4: neuronFeature}
	\vspace{-.4cm}
\end{figure*} 

Another relevant fact is that path based models~\cite{basu2011path2path, batabyal2018elasticpath2path, kanari2018topological} integrate both global (overall shape based approach) and local (vertex or sampled location based approach) features of neuron topology. Topological morphology descriptor (TMD)~\cite{kanari2018topological} aimed at solving the categorization problem, encoded the birth and death of path segments over time in a persistence diagram used as a barcode. The authors showed that TMD exhibits robustness to erroneous 3D sampling and ambiguous branching when the neuron is reconstructed using two different tracing tools. However, the conversion of a discrete 3D reconstructed neuron to the persistence image space is irreversible and many-to-one. Based on the distance used to mark and quantify the birth and death of a branch or component of the neuronal tree, a single persistence image may correspond to multiple neurons. In addition, an appropriate distance measure between persistence diagrams is still unavailable. The work in Path2Path~\cite{basu2011path2path} shows potential to address the neuron cell categorization problem and can be extended to several other related problems, such as synaptogenesis, degeneration of neurons due to neurological diseases, and synaptic plasticity which can be studied by inspecting the path statistics. The work described in this article is motivated by the framework of Path2Path. 

\subsection{What is Path2Path and its variants?}
\label{path2path}
The principle of Path2Path is based on finding the optimal correspondence between the paths of one neuron to that of the other using a proposed metric. It is an intuitive circuit-based approach that appeals to its electrical engineering inventors. In Path2Path, each sampled location on a path is endowed with 3D coordinate values and two features, \textit{concurrence} and \textit{hierarchy}. The concurrence value at each location denotes the number of paths from the soma to dendritic ends that visit that node. The hierarchy value at a location indicates the depth of the location from the soma in terms of the number of bifurcations between the point and the soma. The hierarchy value of a location counts the number of bifurcations one has to cross while traversing from the soma to that location. Using the 3D coordinates, concurrence, and hierarchy values of each location on a path, authors in Path2Path proposed an empirical metric that outputs a distance value between two paths. A path from a neuron corresponds to a path from another neuron if the distance between the paths is minimum over all the paths of the latter neuron. 

This approach has several drawbacks. The Path2Path algorithm is dependent on the number of sampled locations of each path and the registration. The selection of the metric is arbitrary in a sense that the metric is null when two paths have the same set of concurrence values but different locations and hierarchy values. Therefore, it does not qualify the axioms of a metric. In addition, the proposed distance measure uses the Euclidean distance between two paths as a part of the distance computation routine, which favors the pair if they are aligned in proximity after registration. The algorithm of finding the correspondence is not one-to-one and it often leads to the degenerate case where all paths from one neuron are matched to only one path in the second neuron. The problem exacerbates when the number of samples in the two paths are unequal. One potential solution is to resample each path using a constant step~\cite{wan2015blastneuron}, but may, unfortunately, eliminate the importance of the locations, such as curvature of a rooted path prior to resampling. 
\begin{figure*}[t]
\vspace{-.2cm}	
	\centering
	\includegraphics[width=16.6cm, height=9.5cm]{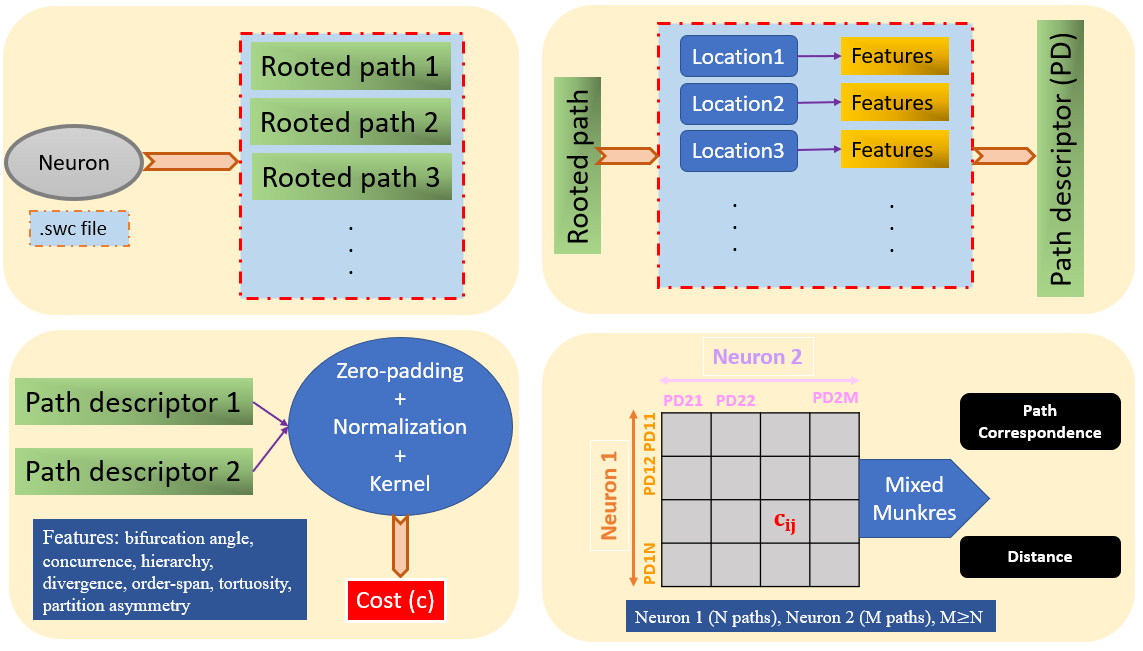}
	\caption{The figure presents a schematic representation of our proposed method and work flow. \textbf{[Top Left]:}~ An SWC~\cite{stockley1993system} file encoding the 3D reconstruction of a neuron is read, and later the neuron is decomposed into an assembly of rooted paths. Each rooted path, spanning from the soma to a synaptic terminal, contains the 3D coordinates of each location traced on the path. \textbf{[Top Right]:}~Each rooted path is subjected to feature extraction from each location on the path. The exhaustive list of the features that are used in our approach is given in the bottom left (blue box). We extract $7$ features, implying that the path descriptor is a matrix of dimension $(\text{number-of-samples}\times 7)$.   }
	\label{fig55: algorithm block}
	\vspace{-.3cm}
\end{figure*} 
ElasticPath2Path~\cite{batabyal2018elasticpath2path} attempted to address the previously mentioned problems. It introduced a mid-point based resampling routine as opposed to constant-length resampling. To ensure one-to-one correspondence between a pair of paths from two different neurons, the Munkres algorithm~\cite{munkres1957algorithms} is employed. Most importantly, elasticPath2Path envisaged the problem of distinguishing two neurons as a continuous deformation between the corresponding paths of the neurons. Such homeomorphism is computed by applying the square root velocity function (SRVF)~\cite{srivastava2011shape} to the Euclidean coordinates of each sample on a path.             
The visual deformation of the corresponding paths has an enormous impact in the validation of the path based on customized features and the proposed distance measure. 
On the flip side, elasticPath2Path failed to address the problem where there is a significant difference in the number of paths between two neurons. As the correspondence is one-to-one, it asserts that elasticPath2Path performs subgraph matching. Both Path2Path and elasticPath2Path did not consider important anatomical morphometrics, such as bifurcation angles and partition asymmetry.
\vspace{-0.5cm}
\subsection{Key aspects of NeuroPath2Path}
\vspace{-0.2cm}
$\boldsymbol{1)}$ The inception of NeuroPath2Path comes from the realization of neuromorphogenesis and the self-similar phenotype of neuronal arbor. Since its birth from the soma, a path of a neuron has an \textit{exploratory} attribute to collect external resources by the \textit{minimal-length-maximal-routing}~\cite{sporns2005human} strategy. Due to the parsimonious exploitation of intrinsic resources (ion density, ATP and other electrophysiological items), the path, which fails to procure external resources, retracts. The exploratory attribute of a path can be expressed by the concurrence values at each sample point of a path. More paths imply more exploration. As the path matures, it has a \textit{competitive} attribute~\cite{miina2002application, genet2014incorporating, lopez2011models} with respect to the other paths in its neighborhood in order to form a synapse. To account for competition, we count the number of paths in the proximity of each sampled location on a path and assign the count to that location. 

$\boldsymbol{2)}$ The fractal dimension~\cite{puvskavs2015fractal, brown2008quantifying} of a neuronal arbor is considered one of the key morphometrics because the fan-out branches of a neuron bear self-similarity. In Path2Path and elasticPath2Path, the notion of matching the paths ignores this important feature. We extend the use of Munkres algorithm to perform one-to-one matching in a sequential fashion, which replicates the self-similar behavior.

$\boldsymbol{3)}$ As path features, we consider the bifurcation angle, partition asymmetry, and fragmentation score to each 3D location on a path. It is shown that the distribution of bifurcation angles in the basal dendrites of cortical pyramidal cells follows a Von Mises distribution~\cite{bielza2014branching}. An experimentally observed fact is that the mean bifurcation angle of branches ordered in a reversed fashion is discriminative for pyramidal cells in different cortical regions. However, the mean bifurcation angle of branches in standard order remains similar for the pyramidal cells. We take the standard ordering of branches, instead of the reverse order, to discriminate different neuronal cell types. Partition asymmetry~\cite{brown2008quantifying, polavaram2014statistical,samsonovich2006morphological} is another visually-significant morphometric. We use the \textit{caulescence} measure as defined in~\cite{brown2008quantifying} to account for the tree asymmetry.

$\boldsymbol{4)}$ We provide visualization of the continuous deformation between a pair of neurons and enumerate path similarity statistics to justify the correspondences between the paths. In contrast, conventional methods perform feature customization and extraction, and the classification, in supervised or unsupervised settings, depends on the abstract feature space and the strength of the classifier. In those methods, the mapping between the space of 3D reconstructed neurons and the feature space is irreversible and abstract. Therefore, apart from the statistical quantification and analysis, it is ambiguous whether improved accuracy of the categorization stems from the trained classifier or the discriminating strength of the extracted features or both. In NeuroPath2Path, the classification problem is modeled as a variant of the transport problem. First, the correspondence of paths between a pair of neurons are decided in the feature space. Next, the correspondence is utilized to deform one neuron to the other. The distances computed between the paths and the deformation together justify the validity of the correspondence.
\begin{figure}[t]
\vspace{-.1cm}	
	\centering
	\includegraphics[width=8cm, height=4cm]{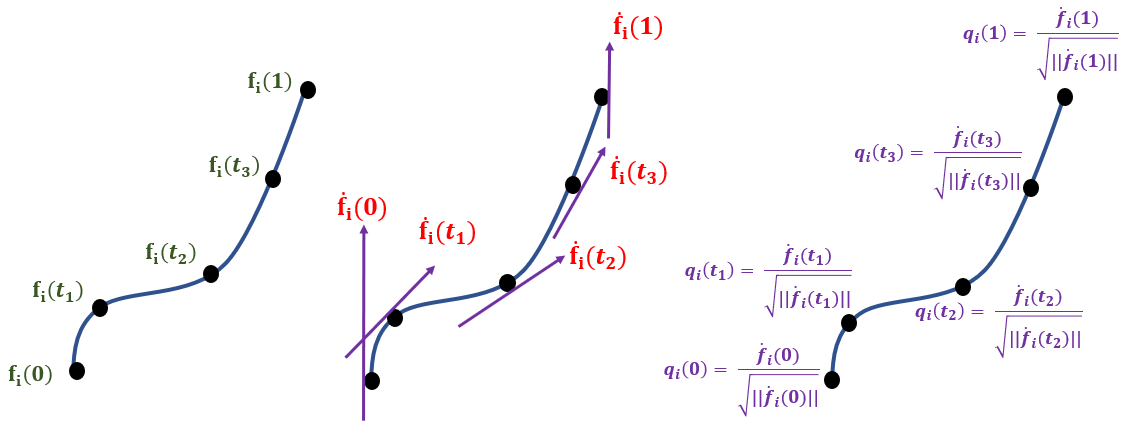}
	\caption{A schematic representation of the square root velocity function ($q$), which is computed at locations on an open curve. This function endows the curve with elasticity so that it can continuously deform (bend, stretch, shrink) to another curve. In a neuron, each rooted path can be modeled as an elastic open curve. }
	\label{fig12: SRVF}
	\vspace{-.5cm}
\end{figure} 

$\boldsymbol{5)}$ With suitable feature selection, NeuroP2P framework can be applied to perform morphological analysis of any cell type with ramified branching arbors, such as microglia and astrocytes. The continuum that is present in the evolution from one cell type to the other can be utilized in the analysis of cell differentiation. As an example, under certain constraints, the strategy of continuous morphing with branch splitting (explained later) can retrieve the intermediate states of a neuron cell while it evolves from a neural progenitor cell to its fully developed state. In short, NeuroP2P can serve as an effective tool for cell-specific informatics, which is not restricted to classification only.

\section{Path modeling of a neuron}
\label{pathmodel}
As mentioned in the introduction, a digitally-traced 3D sampled neuron can be modeled as a graph. Let the graph be represented by $\mathcal{G}=\big(\mathcal{V}, \mathcal{E}, \mathcal{W}\big)$, where $\mathcal{V}$ is the number of 3D locations as vertices and $\mathcal{E}$ is the set of edges connecting the vertices with the corresponding weights $\mathcal{W}$~\cite{harary1969graph}. $\mathcal{G}$ is said to be simple if it does not contain multiple edges between any two vertices. A graph is called undirected if there is no preferred direction associated to an edge. A sequence of contiguous edges is called a path if no vertex and edge are repeated in that sequence. A path of length $k$ has $k$ number of edges or equivalently $(k+1)$ vertices. A sequence of contiguous edges is called a trail if no edge is repeated. If all the vertices except the start and the end of a trail are distinct, it is called a loop. A simple graph without a loop is termed as a tree. If the degree of each vertex is fixed, tree has the fastest growth by volume, hence smallest curvature~\cite{lin2010ricci}. A graph is said to be single-connected if there exists at least one path between a pair of vertices. 
In case of a neuron, $\mathcal{G}$ is a \textit{simple, undirected, weighted,} and \textit{single-connected} tree. 

A path can be considered as an open curve, $f_i(t),~t\in[0,1]$, as defined in differential geometry. The cardinality of the set of vertices, or ,equivalently, the total number of 3D locations, is given by $|\mathcal{V}| = N$. Here, there are $n$ dendritic terminals, which implies that the total number of paths rooted at the soma is $n$. Let $\Gamma$ be the set containing all the paths $f_i,~i\in\{1,2,..,n\}$, which is a linear subspace of the classical Wiener space. Each path is sampled with the number of samples as $\phi$ with the sampled path denoted by $\tilde{f}_i$. We extract $K$ features for each sample on $\tilde{f}_i$, which can be compactly given by the feature matrix for $f_i$ as $\Theta_i\in\mathbb{R}^{\phi\times K}$. Let $\Theta$ be the ordered set containing the feature matrix for all the paths, $\Theta = \{\Theta_1,\Theta_2,...,\Theta_n\}$, where $\Theta_i$ corresponds to the $i^{th}$ path $\tilde{f}_i$.
The \textit{path model} of a neuron $\mathcal{G}$ can be mathematically represented as $H = \{\Gamma, \Theta, \mu \}$, where $\mu$ is a measure that we define in the next section. Note that we use the set of paths, $\Gamma$, as an ordered set which has a one-to-one correspondence with the elements in $\Theta$.   
The standard branch order of a path, $f_i$, is defined as the order in which the locations of bifurcation on a path are visited from the root to the end of the path. Similarly, the reverse branch order is defined when the direction of traversal is reversed. For interclass comparison of neurons, we use the standard order. Whereas, for intraclass comparison, we follow the reverse order.

\section{Proposed methodology}
Our proposed method, which is sequential, scalable and modular, consists of four key stages as depicted in Fig.~\ref{fig55: algorithm block}. In the first stage, centrally curated files of 3D-traced neurons in SWC format (or equivalent formats) are read and then preprocessed to extract only the dendritic arbors, including the soma. Several preprocessing modules, such as range-wise calibration, bifurcation location determination, and synaptic tip identification are employed to aid in preparing an assembly of rooted (soma) paths.

In the second stage, a set of features are extracted from each path, forming a feature descriptor of the path, $\Theta_i\forall i$. An exhaustive list of the features that are used in our method is provided in Section~\ref{featureExtract}, and the systematic quantification of the features is provided in the Appendix. Notice that each path descriptor can be populated with additional structural and geometric features in order to perform fine-grained analysis. 

The central aspect of the third stage is finding an appropriate cost function, as illustrated in Section~\ref{pathdistance}. The cost function assimilates several anatomical features (such as segment length and bifurcation angle) and physiologically relevant factors (such as the competitive behavior, decaying anatomical importance of a path from the soma to synapse). 
A rigorous optimization framework is also formulated to find the relative contributions of such factors. In short, this stage delivers a distance measure between a pair of paths to the last stage. 

\begin{figure*}[t]
\vspace{-.2cm}	
	\centering
	\renewcommand{\tabcolsep}{0.05cm}
\begin{tabular}{ccccc}
	{\includegraphics[width=3.2cm, height=2.5cm]{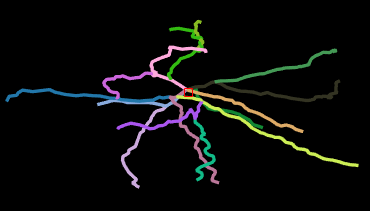}}&
	{\includegraphics[width=3.2cm,height=2.5cm]{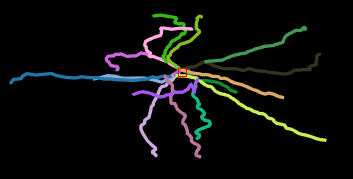}}&
    {\includegraphics[width=3.2cm,height=2.5cm]{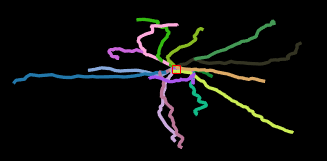}}&
    {\includegraphics[width=3.2cm,height=2.5cm]{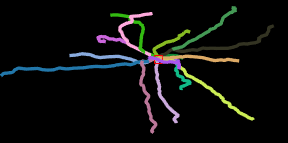}}&
    {\includegraphics[width=3.2cm,height=2.5cm]{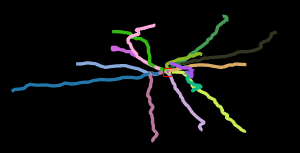}}\\
    {\includegraphics[width=3.2cm,height=2.5cm]{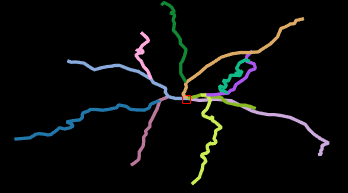}}&
    {\includegraphics[width=3.2cm,height=2.5cm]{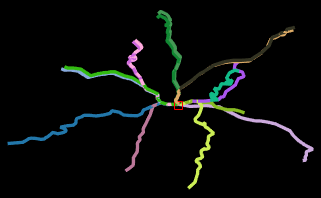}}&
    {\includegraphics[width=3.2cm,height=2.5cm]{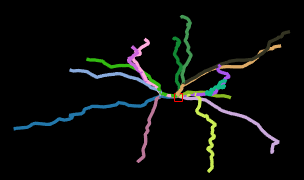}}&
    {\includegraphics[width=3.2cm,height=2.5cm]{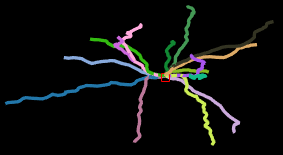}}&
    {\includegraphics[width=3.2cm,height=2.5cm]{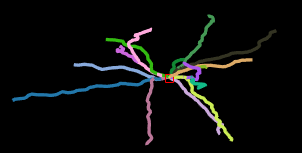}}
\end{tabular}
\caption{The figure depicts the evolution of $15$ paths of one pyramidal neuron to $11$ paths of another pyramidal neuron. Both the neurons are procured and curated from the neocortex (occipital, secondary visual and lateral visual) brain regions of $2$ month-old mice. On the leftmost column, the top figure corresponds to the candidate neuron 1; the bottom figure is the target neuron 2. The evolution is represented in multiple arrays such that the ODD rows are read left-to-right and the EVEN rows are read right-to-left.  
The color associated with each path acts as a marker for the correspondence. At each intermediate step, the morphing of each path is calculated in the SRVF space~\cite{srivastava2011shape} and then the path is projected back in the real 3D domain. In accordance with known properties of the SRVF, the SRVF takes care of the translation between paths. However for visual clarity, we intentionally allow the rotation of each path with respect to the root (soma) while the path advances towards merging with the target path. Prior to applying the SRVF, we reorganize the coordinates of each neuron in decreasing order of the ranges along the $X$, $Y$, and $Z$ axes, implementing an in-place rotation of each neuron.}
	\label{fig: pypymorph}
	\vspace{-.3cm}
\end{figure*}
With a distance measure between neuron paths in hand, this measure is augmented 
in the final stage as elucidated in section~\ref{PathAssign}. We theoretically establish the correspondence of paths of a pair of neurons by repeatedly applying the Munkres algorithm. In contrast to the conventional approaches where the distance between neurons inherently accounts for \textit{sub-graph} matching, we propose a full-tree matching algorithm. 
The repeated application of the Munkres algorithm reveals the fractal or self-similar nature of a pair of neurons. Equivalently, the following question may be posed: how many identical copies, taken together, of the first neuron can match with the second neuron, assuming the second neuron is much larger than the first one? Once the correspondence is found, neurons are diffeomorphically transformed to each other by morphing corresponding paths. This visual representation aids in justifying the correspondence of paths.     

\subsection{Feature extraction on a path}
\label{featureExtract}
We extract a set of discriminating features from each path $f_i\in\Gamma$ of $H$, which are bifurcation angle ($b_i$), concurrence ($C_i$), hierarchy ($\xi_i$), divergence ($\lambda_i$), segment length ($\beta_i$), tortuosity ($\kappa_i$), and partition asymmetry ($\alpha_i$). Therefore, $\Theta_i = [b_i, C_i, \xi_i, \lambda_i, \beta_i, \kappa_i, \alpha_i]\in \mathbb{R}^{\phi\times 7}$.  
Each feature encodes a specific structural property of a neuronal arbor, as described in the Appendix. A schematic of different features along with the systematic quantification is shown in Fig.\ref{fig4: neuronFeature}.

\subsection{Path alignment and path distance measure, $\mu$}
\label{pathdistance}
Given an unequal number of samples in a pair of paths, finding the appropriate distance between two paths or open curves is challenging. Due to the resampling bias imposed by a given tracer, in general, a path contains erroneous sampled locations which could alter the path statistics. For example, adding an extra leaf vertex changes the concurrence values of all the locations on a path. 
 Unlike conventional approaches that used different resampling procedures, such as mid-point based resampling, RANSAC sampling, and spectral sampling, we use the help of the branch order as mentioned in section~\ref{pathmodel} for suboptimal alignment. 
 
 Consider two neurons, $G_1$ and $G_2$, with the corresponding path models given as $H_1$ and $H_2$, respectively. Let $f$ and $g$ be the two paths that are arbitrarily selected from $\Gamma_1$ and $\Gamma_2$, respectively. Without loss of generality, let us assume that $f$ and $g$ contain $\phi_1^b$ and $\phi_2^b$, the number of locations from which the current paths bifurcate. In the case, where $\phi_1^b < \phi_2^b$, we append $(\phi_2^b-\phi_1^b)$ zeros at the end (standard branch order) or at the front (reverse branch order) of a feature vector on $f$.  
 
 Experimental evidence ~\cite{bielza2014branching} suggests that the importance of a bifurcation location on a path decays as one travels the path from the soma to the synaptic end. We utilize this relative importance by way of hierarchy values of the bifurcation locations on a path. Let the sequential order of hierarchy values from the root to the terminal on $f$ be $\xi_f = [\xi_1, \xi_2,...,\xi_{\phi^b_1}]$.  
 Using $\xi_f$, the  $k^{th}$ importance weight is given by $w_k = \frac{1}{\xi_k+\epsilon}/\sum_{j=1}^{\phi^b_1}\frac{1}{\xi_j+\epsilon}$. $\epsilon$ is introduced to avoid the indeterminate case. According to the hierarchy, it is obvious that $\xi_1 < \xi_2 < ... < \xi_{\phi^b_1}$. Thus, $w_1 > w_2>...>w_{\phi^b_1}$. 
 
 Let us consider a feature $\upsilon\in\{b,C,\lambda,\kappa,\beta,\alpha\}$. The values of the feature on the paths, $f$ and $g$, are defined by 
 \begin{eqnarray}
 \upsilon^f &=& [\upsilon_1^f, \upsilon_2^f,...,\upsilon_{\phi_1^b}^f, \underbrace{0,.., 0}_{\phi_2^b-\phi_1^b}]\\\nonumber
 \upsilon^g &=& [\upsilon_1^g, \upsilon_2^g,...,\upsilon_{\phi_1^b}^g] 
\end{eqnarray} 
The distance between $\upsilon^f$ and $\upsilon^g$, weighted by the importance factor, is given by
\begin{eqnarray}
\label{eq:featureDist}
d(\upsilon^{fg}) = \sqrt{\frac{1}{\phi^b_2}\sum_{k=1}^{\phi^b_2}w_k(\upsilon^f_k-\upsilon^g_k)^2}
\end{eqnarray}
This distance is computed for each $\upsilon\in\{b,C,\lambda,\beta,\kappa,\alpha\}$. The overall distance between the paths $f$ and $g$ can be expressed as a weighted average of individual distances.
\begin{eqnarray}
\label{dfg}
\mu^{fg} &=& \delta_1d(b^{fg})+\delta_2d(C^{fg})+\delta_3d(\lambda^{fg})\nonumber\\
        && + \delta_4d(\kappa^{fg}) + \delta_5d(\beta^{fg})+\delta_6d(\alpha^{fg}).
\end{eqnarray}
For simplicity, we take $\delta_i=\frac{1}{6}\forall i$ and consider the final distance as the intrinsic distance between the neurons. For classification, we determine $\delta$ through optimization using $maximizing-interclass-minimizing-intraclass$ distance strategy (See algorithm~\ref{alg2} in the Appendix). We term $\delta$ as the relative importance of features.

\subsection{Path assignment and self-similarity}
\label{PathAssign}
Let the number of paths in $H_1$ be $|\Gamma_1| = n_1$. Similarly, for $H_2$, this value is $|\Gamma_2|=n_2$. Without loss of generality, let us assume $n_1 \le n_2$. 
Using eq.~\ref{dfg}, the cost matrix of paths between $G_1$ and $G_2$ becomes $\mathcal{D}$ ($\mathcal{D}_{ij} = \mu^{ij},~i\in\Gamma_1,j\in\Gamma_2$).
By applying an analogy for the path assignment as a job assignment problem with $n_1$ workers and $n_2$ jobs, we adopt the Munkres algorithm to find the optimal assignment of jobs to the workers from $\mathcal{D}$. In most cases, including inter- and intra-cellular neurons, the job assignment problem is an unbalanced $n_1 < n_2$. We append $(n_2-n_1)$ zero rows to $\mathcal{D}$ to serve as dummy workers. ElasticPath2Path~\cite{batabyal2018elasticpath2path} employed this technique and resulted in an output of $n_1$ optimally matched paths between $G_1$ and $G_2$. However, this is essentially subgraph matching, which may lead to misclassification while dealing with two structurally similar, but different, cell types. For example, hippocampal CA3 pyramidal and cerebellar Purkinje cells have similar dendritic branch patterns, but significantly different number of paths.
To resolve this problem we devise an algorithm, given in the Appendix, by applying Munkres algorithm repeatedly to obtain a full-tree matching. To meet such criterion, the algorithm gives $n_2$ pair of paths. Let the pair be $(\gamma_{11},\gamma_{21}),...,(\gamma_{1n_2},\gamma_{2n_2})$, where $\gamma_{1i}\in \Gamma_1$ and $\gamma_{2j}\in\Gamma_2$. Recall that $n_1<n_2$, which implies that some of the $\gamma_{1i}$ are repeated while forming the pair. Finally, the distance between $G_1$ an $G_2$ is given by 
\begin{eqnarray}
\chi_{G_1G_2} = \sum_{k=1}^{n_2}\mu^{\gamma_{1k}\gamma_{2k}}
\end{eqnarray}

Let $\lfloor\frac{n2}{n1}\rfloor = T$. Then, this procedure to find the correspondence is termed as $T-$regular matching, which in turn can be thought of $T$ nearly self-similar structures akin to a fractal system. The detailed algorithm is provided in the Appendix. 
\begin{figure}[h]
\vspace{.1cm}	
	\centering
	\renewcommand{\tabcolsep}{0.01cm}
\begin{tabular}{cc}
	{\includegraphics[width=4.5cm, height=3cm]{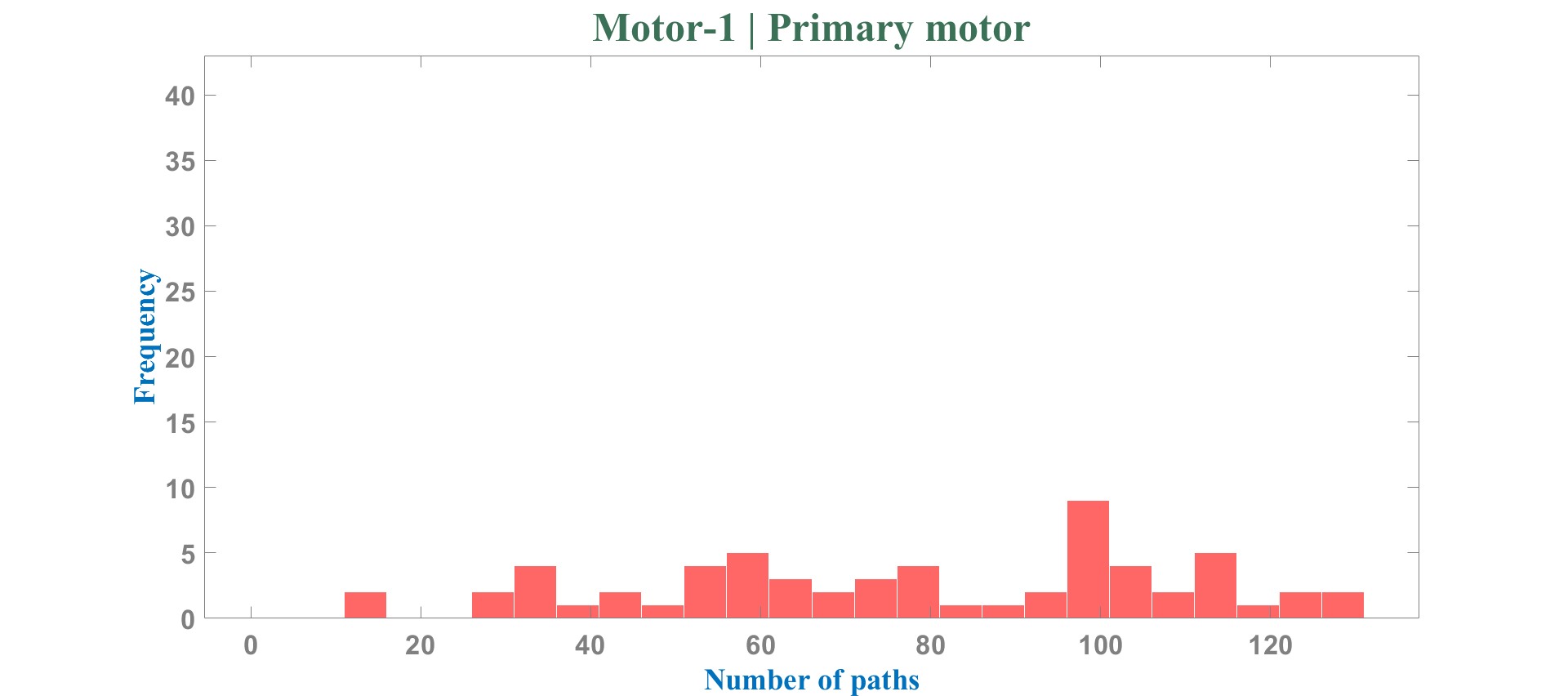}}&
	{\includegraphics[width=4.5cm,height=3cm]{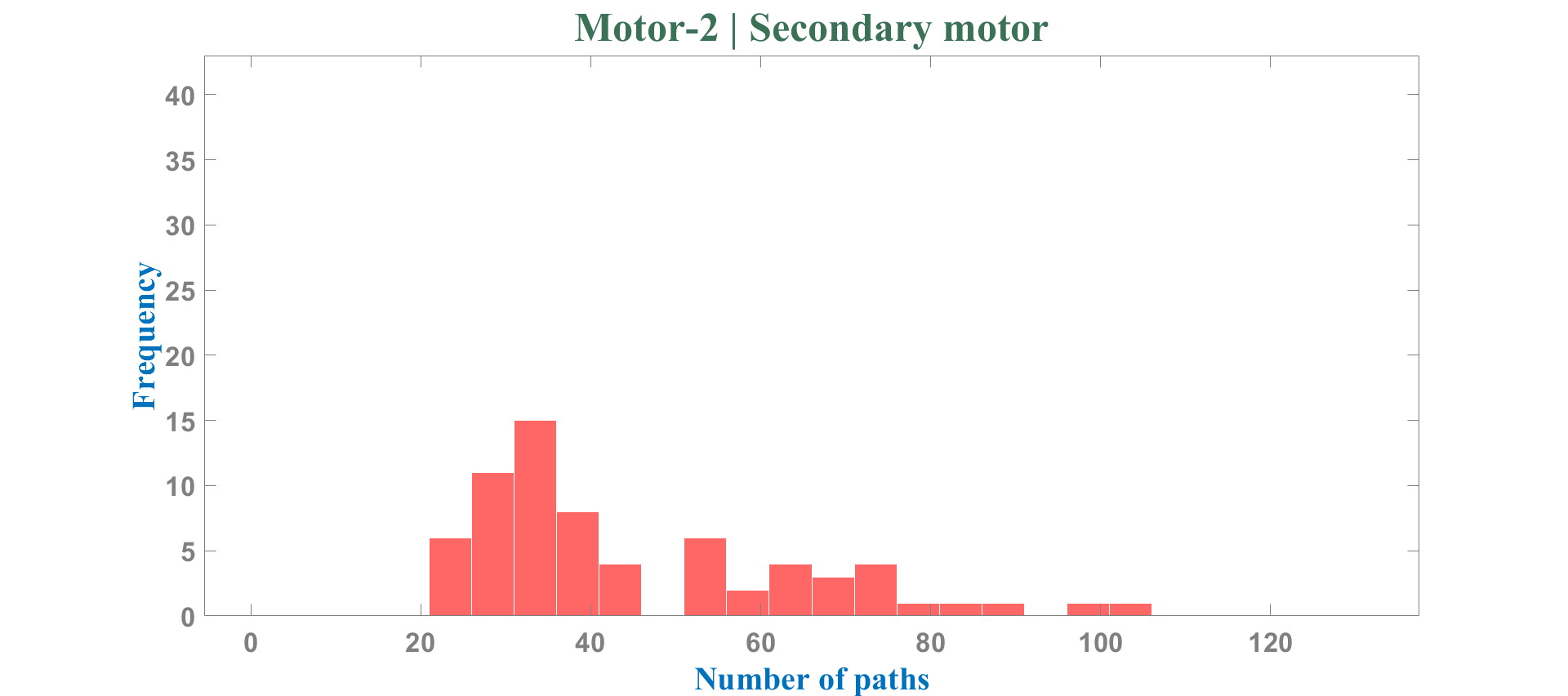}}\\
    {\includegraphics[width=4.5cm,height=3cm]{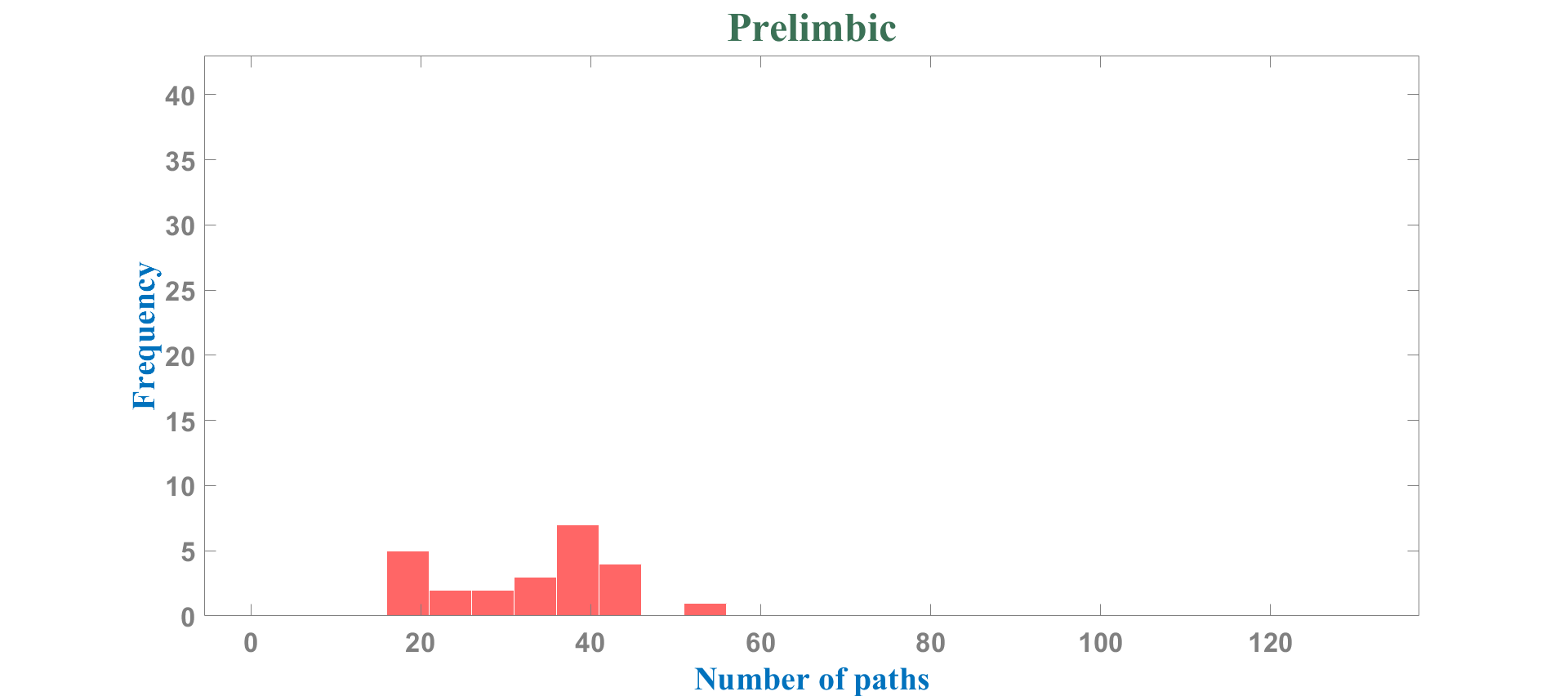}}&
    {\includegraphics[width=4.5cm,height=3cm]{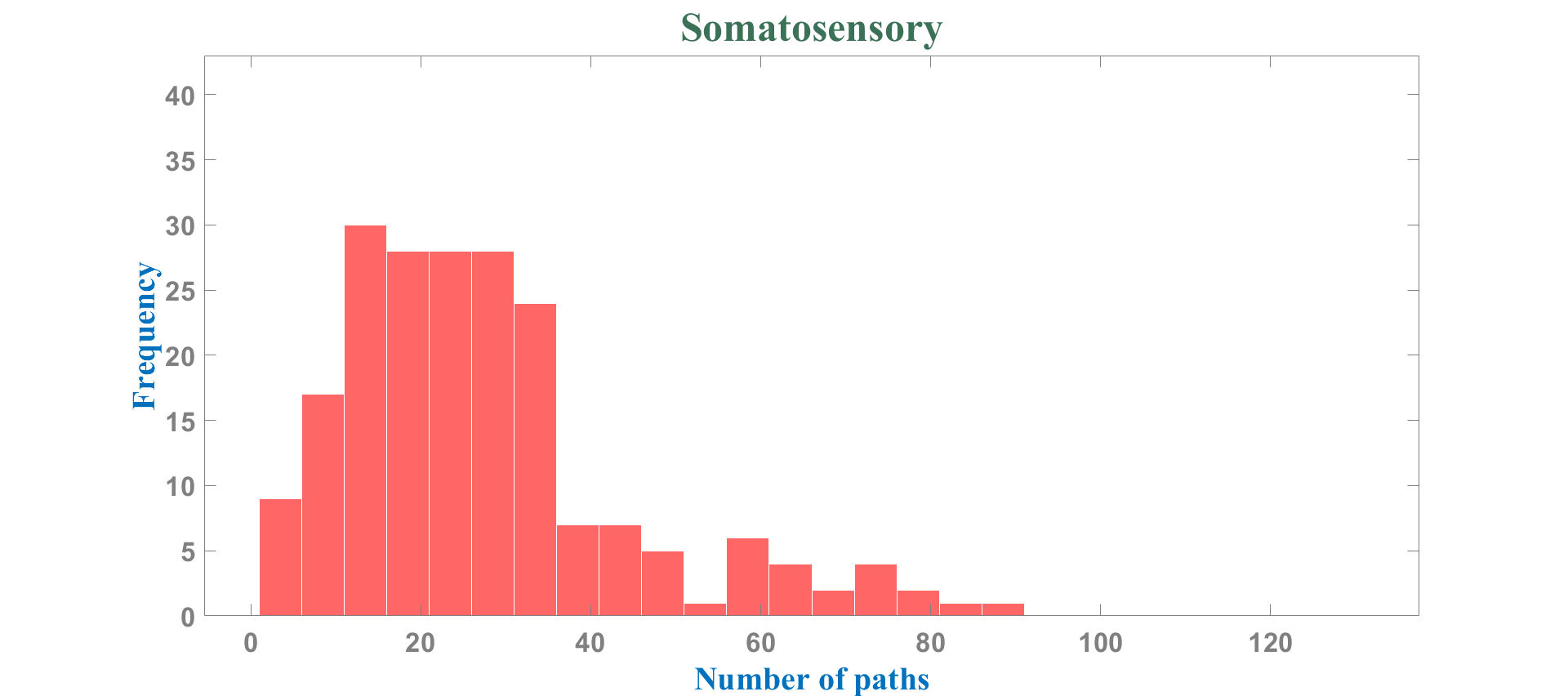}}\\
    {\includegraphics[width=4.5cm,height=3cm]{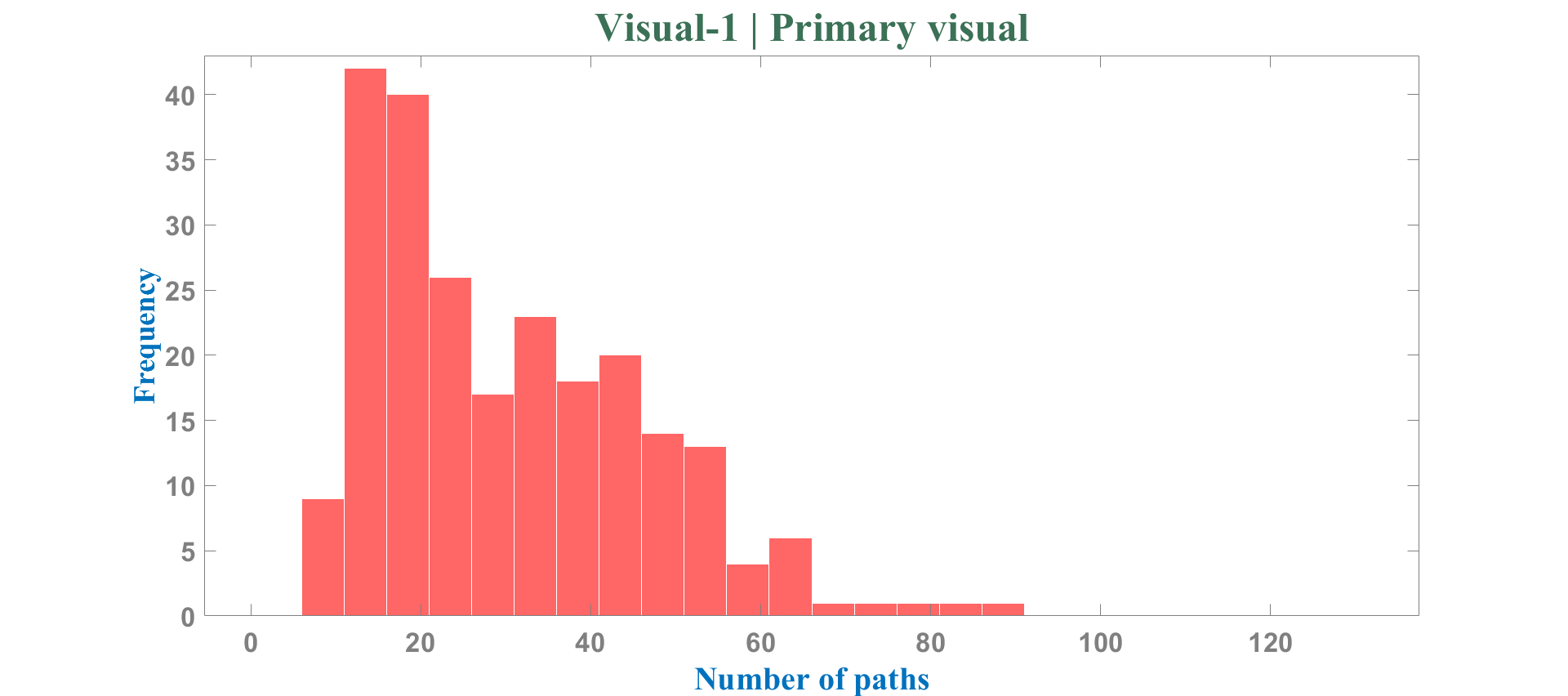}}&
    {\includegraphics[width=4.5cm,height=3cm]{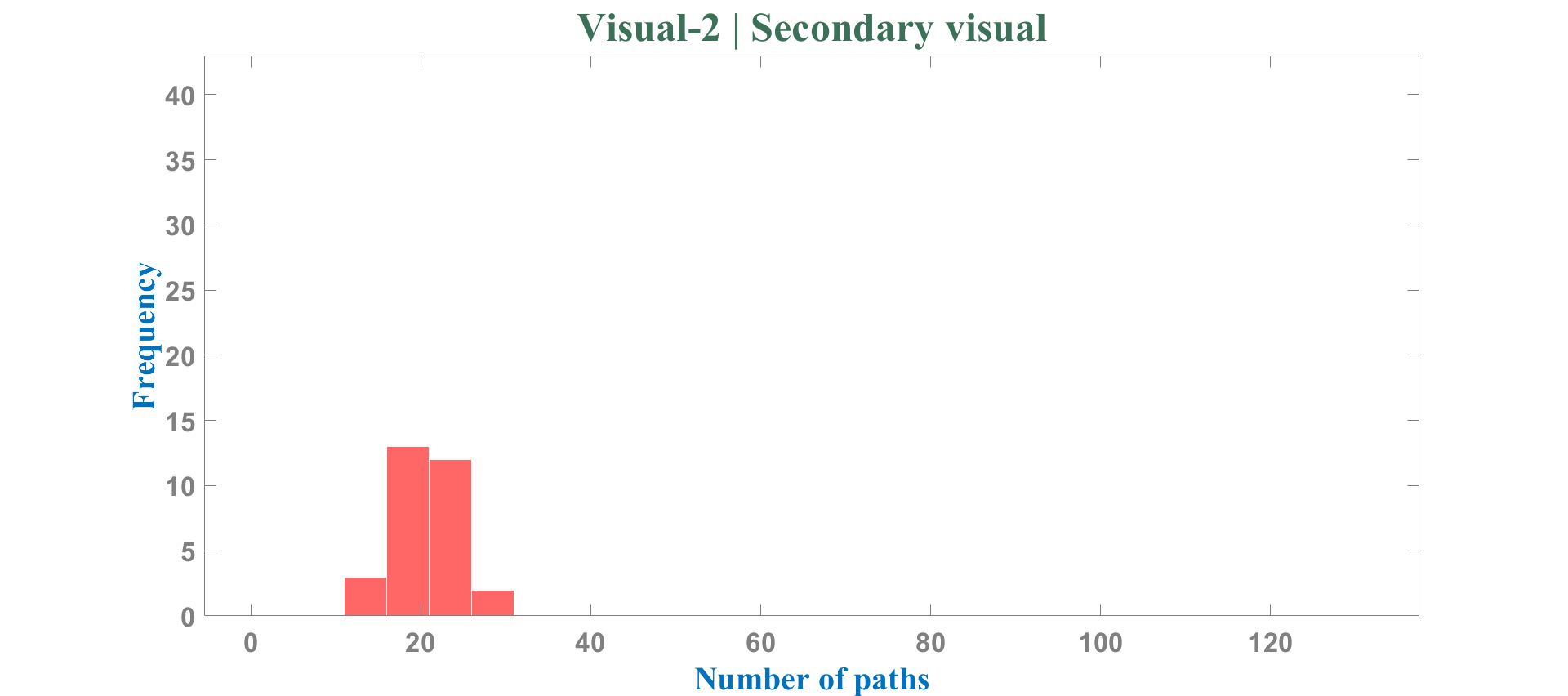}}\\
\end{tabular}
\caption{The figure shows the distribution of paths in each pyramidal cell category from Dataset 1. By glancing at the distribution profiles, a set of inferences can be drawn. The distribution of paths in primary motor is fairly uniform. For neurons from the somatosensory cortex and primary visual cortex, the histogram is right-skewed, indicating a majority of neurons with the number of paths lying in the range $[10,40]$. The probability distribution of somatosensory pyramidal neurons resembles a right-skewed gamma distribution, and that of primary visual neurons closely follows an exponential distribution. The profiles of secondary visual and prelimbic neurons are poorly understood due to scarcity of samples. Most importantly, their distributions are entirely overlapped (within $[10,40]$) in the region where the majority of primary visual and somatosensory neurons can be sampled. From the figure, it is evident that the number of paths alone is not sufficiently discriminatory.}
	\label{fig: pathdist}
	\vspace{-.1cm}
\end{figure}
There are four modules that are sequentially executed in the algorithm. The first module mathematically deciphers the relatively self-similar anatomy of a larger neuron compared to a smaller one, yielding the number of copies of the smaller one needed to stitch together to approximately obtain the larger one. The routine runs for $\lfloor\frac{n_2}{n_1}\rfloor$ times, which indicates that each path in neuron 1 (containing $n_1$ paths), is matched with $\lfloor\frac{n_2}{n_1}\rfloor$ paths of neuron 2 (containing $n_2$ paths). Here $n_2 > n_1$.

The second module runs for the remaining unpaired paths of neuron 2. The assigned correspondence is added to the list of paired paths from the first module. However, not all the pairs are anatomically consistent. This is dictated by an internal constraint of Munkres algorithm, in which the assignment is carried out without replacement. In the Munkres algorithm, if one `worker'(a path from neuron 1) is assigned a `job' (a path from neuron 2), then the 'job' is not available for further assignment. Therefore, if the distance between two paths is significantly large, it demands further inspection whether the pair of paths is morphologically different to each other or the algorithmic constraint induces the large distance value.  
 This motivates us to introduce the third module.   

\begin{figure*}[t]
\vspace{-.2cm}	
	\centering
	\renewcommand{\tabcolsep}{0.05cm}
\begin{tabular}{cc@{\hskip 1.4cm}cc}
	{\includegraphics[width=3.2cm,height=2.8cm]{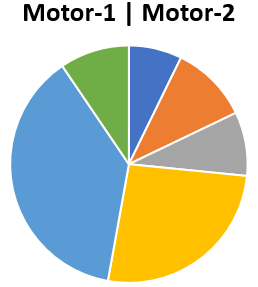}}&
	{\includegraphics[width=3.2cm,height=2.8cm]{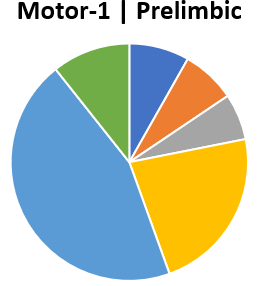}}&
    {\includegraphics[width=3.2cm,height=2.8cm]{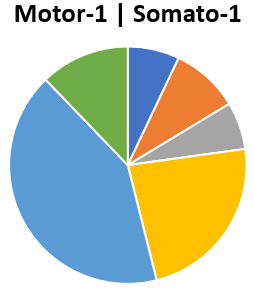}}&
    {\includegraphics[width=3.2cm,height=2.8cm]{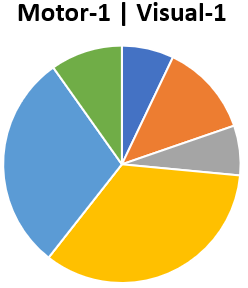}}\\
    {\includegraphics[width=3.2cm,height=2.8cm]{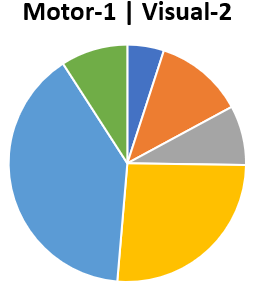}}&
    {\includegraphics[width=3.2cm,height=2.8cm]{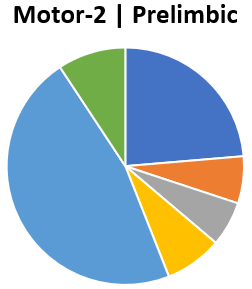}}&
    {\includegraphics[width=3.2cm,height=2.8cm]{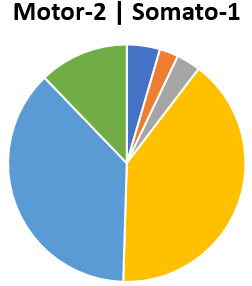}}&
    {\includegraphics[width=3.2cm,height=2.8cm]{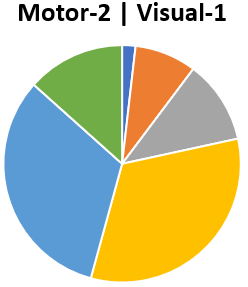}}\\
    {\includegraphics[width=3.2cm,height=2.8cm]{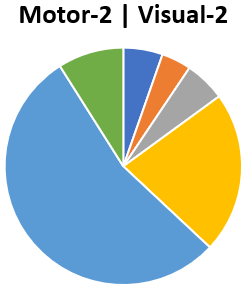}}&
    {\includegraphics[width=3.2cm,height=2.8cm]{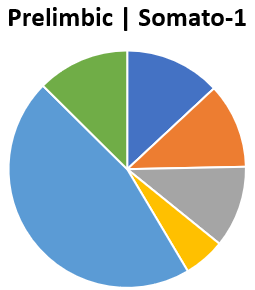}}&
    {\includegraphics[width=3.2cm,height=2.8cm]{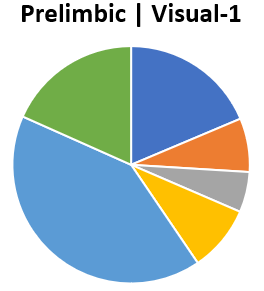}}&
    {\includegraphics[width=3.2cm,height=2.8cm]{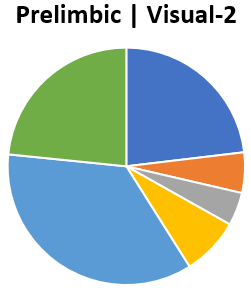}}\\
    {\includegraphics[width=3.2cm,height=2.8cm]{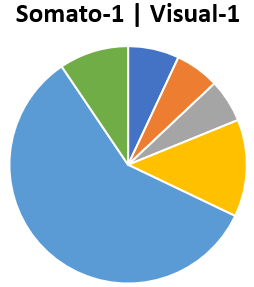}}&
    {\includegraphics[width=3.2cm,height=2.8cm]{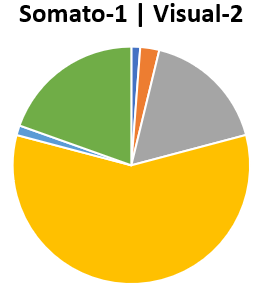}}&
    {\includegraphics[width=3.2cm,height=2.8cm]{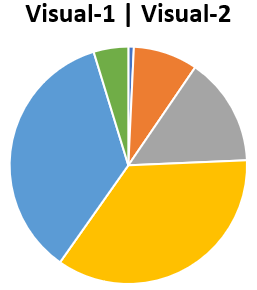}}&
    {\includegraphics[width=3.2cm,height=2.8cm]{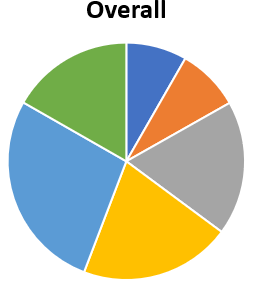}}\\
\end{tabular}
\caption{Relative importance $\delta$ for subsets of classes of Dataset 1. Each color corresponds to a specific feature and the area, subtended by the color in a pie chart, indicates its relative importance. By property, $\sum\delta=1$, which implies a probability distribution. The color codes are as follows. \textbf{green}-divergence, \textbf{orange}-bifurcation angle, \textbf{gray}-partition asymmetry, \textbf{yellow}-concurrence, \textbf{deep blue}-tortuosity, \textbf{sky blue}-segment length. The pie charts taken together asserts a set of inferences. ($1$) The relative importance of features $\delta$ of all the classes (marked `overall') somewhat follows a uniform distribution. ($2$) Segment length and concurrence are two predominant features when the pyramidal neurons from primary motor cortex (motor-1) are compared to the rest of the classes. ($3$) For the prelimbic class, divergence, tortuosity, and segment length appear to be most important. ($4$) $\delta$ for the somatosensory class toggles between two distributions with comparatively smaller and larger importance of concurrence. }
	\label{fig: importance1}
	\vspace{-.4cm}
\end{figure*}

In the third module, we inspect the pair of paths having distances more than a threshold. The threshold is selected based on the skewness, median and standard deviation of the distance values.  
As mentioned earlier, in order to find the distance of a feature on two paths (eq.~\ref{eq:featureDist}), we append zeros to the path having relatively fewer number of locations than the other. The choice of traversal order dictates to which side the zeros are appended. Notice that more zeros lead to higher distance value between paths, and this happens only when there is significant mismatch in the highest level of hierarchy. This fact can be interpreted from the morphological viewpoint. A path with a large number of bifurcation locations (so, large maximum hierarchy value), called a central path of a neuron, exploits the environment of the neuron extensively when compared to path with fewer number of bifurcations. Unless otherwise required, a path with large hierarchy values should not be compared with a path with much smaller maximum hierarchy value.  
The highest level of hierarchy values of two paths are given by $h_1$ and $h_2$ with $h_1<h_2$. We set a criteria that if $|h1-h2| > \frac{max[h1,h2]}{2}$, we do not consider the distance between the pair, and opt for the best match in terms of minimum distance for each path of the pair separately. This is outlined in the reassignment module. The reassigned pairs are added to the list of paired paths serving as the list of correspondence. 

\subsection{Path morphing}
\label{elasticMorph}
Once the correspondence of paths between neurons is established, it is imperative to know the structural similarity between the paths - \textit{whether a pair of paths are structurally similar to each other, or the pair is structurally incoherent but the algorithm outputs such a pair due to its internal constraints.} 
This is achieved in two ways: with a visual representation by morphing the paths of one neuron to that of the other using an elastic framework, and by extracting path statistics.
\begin{figure*}[t]
\vspace{-.1cm}	
	\centering
	\renewcommand{\tabcolsep}{0.05cm}
\begin{tabular}{ccccc}
	{\includegraphics[width=3.2cm, height=2.5cm]{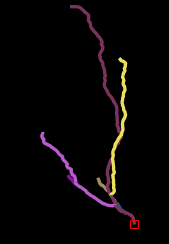}}&
	{\includegraphics[width=3.2cm,height=2.5cm]{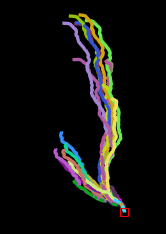}}&
    {\includegraphics[width=3.2cm,height=2.5cm]{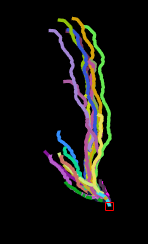}}&
    {\includegraphics[width=3.2cm,height=2.5cm]{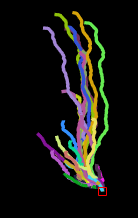}}&
    {\includegraphics[width=3.2cm,height=2.5cm]{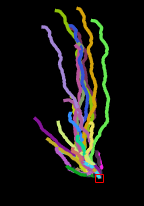}}\\
    {\includegraphics[width=3.2cm,height=2.5cm]{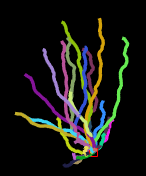}}&
    {\includegraphics[width=3.2cm,height=2.5cm]{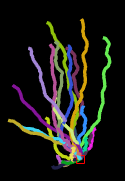}}&
    {\includegraphics[width=3.2cm,height=2.5cm]{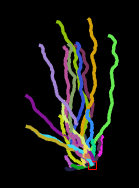}}&
    {\includegraphics[width=3.2cm,height=2.5cm]{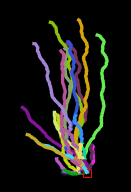}}&
    {\includegraphics[width=3.2cm,height=2.5cm]{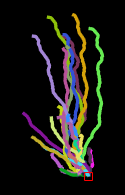}}\\
    {\includegraphics[width=3.2cm,height=2.5cm]{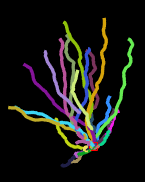}}&
    {\includegraphics[width=3.2cm,height=2.5cm]{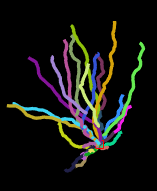}}&
    {\includegraphics[width=3.2cm,height=2.5cm]{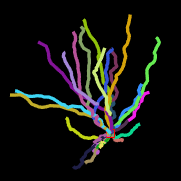}}&
    {\includegraphics[width=3.2cm,height=2.5cm]{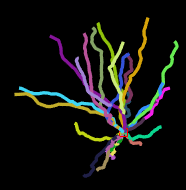}}&
    {\includegraphics[width=3.2cm,height=2.5cm]{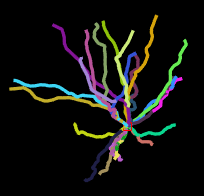}}\\
    {\includegraphics[width=3.2cm,height=2.5cm]{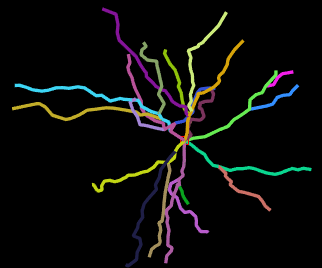}}&
    {\includegraphics[width=3.2cm,height=2.5cm]{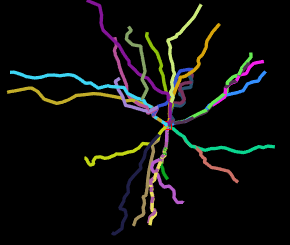}}&
    {\includegraphics[width=3.2cm,height=2.5cm]{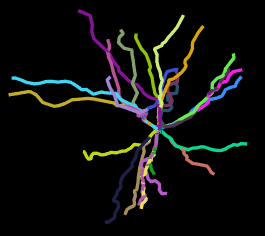}}&
    {\includegraphics[width=3.2cm,height=2.5cm]{Gr2Py37.PNG}}&
    {\includegraphics[width=3.2cm,height=2.5cm]{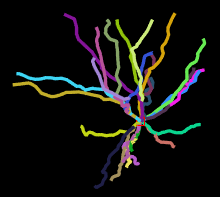}}\\
\end{tabular}
\caption{This gallery of images captures the progressive evolution of paths from a granule neuron to a pyramidal one. The granule neuron is procured from the hippocampus (dentate gyrus) of a $5$ month-old mouse, containing $6$ rooted paths. The pyramidal neuron is sampled from the neocortex (occipital lobe, secondary visual, lateral visual) of a $2$ months old mouse, containing $22$ paths. The evolution is represented in multiple arrays such that the ODD rows are read left-to-right and the EVEN rows are read right-to-left. In the first column, the top image is the granule cell and at the bottom is the pyramidal one. Structurally, the pyramidal neuron is larger than the granule one. However, they are properly scaled to fit for visualization.}
	\label{fig: grpymorph}
	\vspace{-.4cm}
\end{figure*}
A rooted path of a neuron can be considered as an open curve as shown in Fig.~\ref{fig12: SRVF}\cite{batabyal2018elasticpath2path,srivastava2011shape}. Each location on the path can be considered as a function of a parameter,$t\in[0,1]$. The square root velocity function (SRVF) that is applied on a location $f(t)$ is defined as $q(t) = \frac{\dot{f}(t)}{\sqrt{||\dot{f}(t)||}}$. For a pair of paths $i$ and $j$, we obtain $q_i$ and $q_j$, which assists in retrieving the intermediate deformations as linear combinations of $q_i$ and $q_j$ given by $q_{ij}^n = q_i(1-n) + nq_j;~n\in[0,1]$. $n$ denotes the intermediate algorithmic time steps. Although the deformations are exhibited using the 3D coordinates of the locations of a path, the deformations can also be computed in the feature domain. 
An example of the continuous morphing process between two pyramidal neurons from the secondary visual cortex of the mouse is shown in Fig.~\ref{fig: pypymorph}. The $15$ paths of the former neuron merge with $11$ paths of the latter upon termination of the morphing process. This implies that more than one path of the first neuron have the same final destination path of the second neuron. It is noted that our algorithm does not consider the costs that are incurred by the merging or splitting of paths during progression. The assessment of such costs requires biophysical measurements of neurons, such as metabolic cost of merging or splitting of branches. Therefore, the cost between paths in eq.~\ref{dfg} is proportional to the cost of structural disparity instead of biophysical costs.    

The prime question is: why do we need to inspect intermediate deformations? Statistical assessment of anatomical similarities between paths is \textit{sufficient} to validate the correspondence that is obtained from the Munkres algorithm. However, to make the correspondence \textit{necessary}, the intermediate deformations should comply with key cell-type characteristics~\cite{srivastava2011shape}. So we use the SRVF framework to show the deformations so that any noticeable incoherence can be attributed to the feature selection, distance measurement, or both algorithms even though we might obtain improved classification accuracy in the end. 
\vspace{-0.6cm}
\section{Datasets and results}
\vspace{-0.2cm}
We validate the approach on two datasets that are collected from a centrally curated on-line repository of 3D reconstructed neurons,  Neuromorpho.org~\cite{ascoli2007neuromorpho}. To demonstrate the strength of our approach, one dataset is compiled for intraclass and the other one for interclass analysis and comparison. 
\vspace{-0.4cm}
\subsection{Dataset-1 (Intraclass)}
\vspace{-0.2cm}
This dataset contains 3D-traced neurons from $6$ distinct regions of the mouse neocortex.
The regions with their cortical locations are visual-1 or primary visual (occipital), visual-2 or secondary visual (occipital), prelimbic (prefrontal), somato-1 or primary somatosensory (somatosensory), motor-1 or primary motor (frontal), and motor-2 or secondary motor (frontal).  

We experiment with $62$ neurons of motor-1, $68$ neurons of motor-2, $24$ neurons of prelimbic, $204$ neurons of somato-1, $237$ neurons of visual-1, and $30$ visual-2 neurons with $625$ neurons in total. The neurons vary widely in their morphological characteristics, such as the number of paths in each neuron. The histogram of paths corresponding to each category is shown in Fig.~\ref{fig: pathdist}.

Next, we investigate the relative importance of each feature (mentioned in section~\ref{featureExtract}) in terms of $\delta$ when comparing a set of classes. For space constraint, we provide $\delta$ values separately for each pair of classes and all the classes taken together. The relative importance is listed in Fig.~\ref{fig: importance1} by a pool of pie charts. A set of class-specific inferences regarding the relative importance is enlisted in the figure description. Whereas the pie charts present a comprehensive view of feature strength. In practice, however, the values are required to report the distance between a pair of neurons. The values are reported in Table~\ref{importSet1}. 

\begin{table*}[ht]
\vspace{-.1cm}	
	\caption{Importance weight $\delta$ values for dataset-1. For space constraint, we provide feature-specific importance weight for classification in case of pairwise classes and all classes separately. }
	\label{importSet1}
	\centering
	\begin{tabular}{|l!{\vrule width 1.5pt}c|c|c|c|c|c|}
		\hline
		 & Tortuo$(\kappa)$ & Bifur-angle$(b)$ & Part-aym$(\alpha)$ &Concur$(C)$ & Seg-len$(\beta)$ & Diverg$(\lambda)$  \\\hline 
		Motor1-Motor2   &0.0729  &0.1072  &0.0876   &0.2644    &0.3794    &0.0955   \\\hline
		Motor1-Prelimbic &0.0820   &0.0737  &0.0631   &0.2262    &0.4483    &0.1067   \\\hline
		Motor1-Somato   &0.0706  &0.0929  &0.0646   &0.2328    &0.4175    &0.1218   \\\hline
		Motor1-Visual1  &0.0689  &0.1230   &0.0659   &0.3317    &0.2878    &0.1227   \\\hline
		Motor1-Visual2  &0.0499  &0.1212  &0.0812   &0.2612    &0.3952    &0.0912   \\\hline
		Motor2-Prelimbic &0.2719 &0.0737  &0.0703   &0.0899    &0.5367    &0.0484   \\\hline
		Motor2-Somato   &0.0452  &0.0256  &0.0328   &0.4008    &0.3730    &0.1266   \\\hline
		Motor2-Visual1  &0.0167  &0.0773  &0.1041   &0.2996    &0.2969    &0.2055   \\\hline 
		Motor2-Visual2  &0.0545  &0.0411  &0.0564   &0.2235    &0.5489    &0.0756   \\\hline 
		Prelimbic-Somato &0.1314 &0.1168  &0.1121   &0.0568    &0.4621    &0.1187   \\\hline 
		Prelimbic-Visual1 &0.2091 &0.0821 &0.0617   &0.1012    &0.4615    &0.0795   \\\hline 
		Prelimbic-Visual2 &0.2311 &0.0551 &0.0453   &0.0790    &0.3556    &0.2339   \\\hline 
		Somato-Visual1  &0.0693  &0.0598  &0.0584   &0.1319    &0.5816    &0.0941   \\\hline 
		Somato-Visual2  &0.0144  &0.0310  &0.2043   &0.6963    &0.0159    &0.0382   \\\hline
		Visual1-Visual2 &0.0075  &0.0880  &0.1482   &0.3551    &0.3558    &0.0473    \\\hline 
		Overall         &0.0785  &0.0807  &0.1736   &0.1956    &0.2600    &0.2076   \\\hline 
	\end{tabular}
	\vspace{-.3cm}
\end{table*} 

It is worthwhile to note that although there is significant variance in feature strength when all pairs are considered, the distribution approximates a uniform distribution when all classes are taken. This result endorses the selection of our features for all-class classification tasks. It is also important to mention that this framework can incorporate any set of path-specific features, not restricted to our selected features only.  

In order to verify the consistency of the path correspondence obtained from Munkres algorithm (provided in section~\ref{PathAssign}), we statistically evaluate each pair of paths in the correspondence list using pyramidal neurons from two different regions (the somatosensory cortex and secondary visual cortex.) The neuron from the somatosensory cortex (neuron-2) contains $28$ rooted paths, while the other (neuron-1) has $11$ rooted paths. Table~\ref{pathCorrespondence} provides the exhaustive list of path correspondences, distances between corresponding paths, correspondences obtained by a competitive approach named ElasticPath2Path~\cite{batabyal2018elasticpath2path}, and the best correspondences of the paths of neuron-2 with that of neuron-1. Notice that the best correspondence of a path $f$ of neuron-2 is the path $g$ of neuron-1, which yields minimum distance with $f$. Whereas, the Munkres algorithm works on the criteria where the sum of path distances (in our case $11$ paths at a time) is minimized.  

In Table~\ref{pathCorrespondence}, the two columns on the left enumerate the pair that consists of the path number of neuron-1 and that of neuron-2. A subset of paths of neuron-1 is repeated because neuron-2 (with $28$) has more paths than neuron-1 (with $11$). So from neuron-1 to neuron-2, the correspondence is a surjective mapping. This is in contrast with ElasticPath2Path, where the mapping is bijective and, as result of that, the algorithm outputs only $11$ pairs in the correspondence list. The rest of the $28-11=17$ paths are left unmatched, yielding a solution of the subgraph matching problem. The unmatched paths are marked with `NA' in the fourth column.

The last column, tagged as the best match, identifies only $\{4,5,8,9\}$ path indices out of $11$ paths of neuron-1. Nevertheless, this best matching algorithm also elicits a potential solution for subgraph matching. There are certain extreme cases where all paths of one neuron are matched with only one path of the other neuron, posing \textit{degenerate} solutions of the neuron matching problem. We mark the correspondences in yellow, where the results of our algorithm and best match coincide.  

Recall that neuron-1 has $11$ paths and neuron-2 contains $28$ paths. Careful observation of the first column of the table suggests that the set of numbers $\{1,2,...,11\}$ is repeated twice in the serial order followed by $6$ path indices which are $\{9,4,10,8,2,5\}$. Here, Munkres algorithm is applied thrice. Each time Munkres algorithm outputs $11$ pairs of paths for correspondence. Therefore, the first two passes encompass $11*2 = 22$ pairs leaving $28-22=6$ paths of neuron-2 unassigned. Before applying the third pass, the cost matrix $\mathcal{D}$ is cropped with a dimension $\mathcal{R}^{11\times 6}$. The cropped cost matrix is then transposed ($\mathcal{R}^{6\times11}$), zero-appended ($\mathcal{R}^{11\times 11}$) and subjected to Munkres.  
The above observation also indicates that $2$ self-similar copies of neuron-1 approximates neuron-2 in the sense of minimum path to path distance. Therefore, the relative fractal index of neuron-2 with respect to neuron-1 is $2\frac{6}{11}$ or $2.545$.   

The question is: can the arithmetic average (which is $0.92$) of the `Distance' column of Table~\ref{pathCorrespondence} be regarded as the final distance between the neuron-1 and neuron-2? Unfortunately, it is not. The reason is explained in section~\ref{PathAssign} and reiterated briefly in the following sentences. After computing the correspondences (column-1 and column-2), we identify the defective sets of pairs for which there are significant differences in the hierarchy levels. The larger the difference, the larger the number of zeros that are appended to each feature on the path, raising the chances of technical error in the final distance value. In the table, the defective pairs are emboldened with blue color. We delete these pairs and replace the correspondences of path indices $18$ and $9$ (neuron-2) with their best matches from neuron-1.
Note that path indices $7$ and $4$ of neuron-1 have already been matched with other paths of neuron-2, which are $7--28$, $4--11$, and $4--12$. Therefore, those paths are not subjected to re-assignment. After inserting the best matches for the path indices $18$ and $9$ (which are $9$ and $5$ from neuron-1 respectively), the corresponding distance values are noted. This is described in the fourth routine, `Reassignment' of algorithm~\ref{alg2}. The final distance between the two neurons turns out as $0.90$ (rounded off). The competitive approach, ElasticP2P produces a distance value of $0.67$, which implies that the two neurons are more similar. 
This is discordant with the fact that the two neurons are sampled from two different regions and have two distinct arbor types. This disagreement can be explained due to subgraph matching nature of ElasticP2P. Neuron-1 with $11$ paths is well-matched with a part of neuron-2. However, the rest $17$ paths of neuron-2 are structurally dissimilar with neuron-1. In this case, our method, NeuroPath2Path performs significantly better in distinguishing two neurons in terms of distance owing to its full-graph matching property.  

\begin{table*}[ht]
\vspace{-.1cm}	
	\caption{Distance and correspondence between paths. The correspondences between the paths of neuron-1 and neuron-2 are enlisted in the first two columns. The numbers in yellow indicate that the correspondence obtained by NeuroP2P matches with the candidates of best correspondence in the sense of minimum distance.
	 The pairs in blue are subjected for further verification because of large differences in the hierarchy values (Routine $4$ in Algorithm $2$). }
	\label{pathCorrespondence}
	\centering
	\begin{tabular}{|c!{\vrule width 1.5pt}c !{\vrule width 1.5pt}c !{\vrule width 1.5pt} c !{\vrule width 1.5pt} c|}
		\hline
		 Index (Neuron-1)  & Index (Neuron-2) & Distance & ElasticP2P~\cite{batabyal2018elasticpath2path} & Best match\\\hline 
		  1      & 25      &1.0595      & 1     & 9\\\hline
		  2      & 2       &0.8031      & 2     & 5 \\\hline
		  3      & 27      &0.4530      & 3     & 5\\\hline
		  \colorbox{yellow}{4}      & 12      &0.7910      & 4     &  \colorbox{yellow}{4}\\\hline
		   \colorbox{yellow}{5}      & 3       &0.5571      & 5     &  \colorbox{yellow}{5}\\\hline
		  6      & 26      &0.6165      & 6     & 5\\\hline
		  7      & 28      &0.5690      & 7     & 5\\\hline
		   \colorbox{yellow}{8}      & 5       &0.5606      & 8     &  \colorbox{yellow}{8}\\\hline
		  \colorbox{yellow}{9}      & 23      &0.6271      & 9     &  \colorbox{yellow}{9}\\\hline
		  10     & 13      &0.7904      & 10    & 4\\\hline
		  11     & 24      &0.6096      & 11   &5\\\hline
		  1      & 7       &1.4997      & NA    & 8\\\hline
		  2      & 1       &0.9902      & NA    & 9\\\hline 
		  3      & 15      &1.1052      & NA    & 5\\\hline
		  4      & 11      &0.9223      & NA    & 9\\\hline
		   \colorbox{yellow}{5}      & 6       &0.5932      & NA    &  \colorbox{yellow}{5}\\\hline
		  6      & 17      &1.7108      & NA    & 9\\\hline
		  \textbf{\textcolor{blue}{7}}      & \textbf{\textcolor{blue}{18}}      &1.2869      & NA   &9\\\hline
		  8      & 20      &0.8704      & NA    & 9\\\hline
		  9      & 4       &0.7708      & NA    & 5\\\hline
		  10     & 14      &0.9186      & NA    & 5\\\hline
		  11     & 10      &1.1796      & NA    & 9\\\hline
		  9      & 8       &0.8504      & NA    & 5\\\hline
		  \textbf{\textcolor{blue}{4}}      & \textbf{\textcolor{blue}{9}}       &1.2279      & NA    & 5\\\hline
		  10     & 16      &1.1810      & NA    & 5\\\hline
		  8      & 19      &1.1658      & NA    & 9\\\hline
		  2      & 21      &1.1522      & NA    & 5\\\hline
		   \colorbox{yellow}{5}      & 22      &0.8931      & NA    &  \colorbox{yellow}{5}\\\hline
	\end{tabular}
	\vspace{-.1cm}
\end{table*}    

For classification, we compute the importance values $\delta$ from (\ref{dfg}) and show them in Table~\ref{importSet1}. The importance values are applied to compute the distance between a pair of neurons. Using our distance function, we resort to the K nearest neighborhood classifier. We randomly partition the dataset into our training and test set using a constant ratio and rerun the experiment $5$ times. The ratio that we maintain is $0.1$ and $0.2$ as train and test datasets. As the number of paths that a neuron has is a distinguishable feature for certain classes, we devise a strategy to test each neuron from the testing dataset. For a neuron with number of paths as $n_P$, we seek candidate neurons from the training set with the number of paths ranging in $[n_P-L, n_P+L]$. Overall, NeuroPath2Path contains two hyperparameters, $K$ (number of nearest neighbors) and $L$. THis step is followed by the identical testing procedure while considering the interclass dataset. We fixed $L=50$ for our experiments. 
As noted before, we adopt the reverse and standard branch orders for dataset-1 and dataset-2, respectively.

With the ratio of train and test as $8:2$, the confusion matrix of NeuroPath2Path for an instance of random partition of data is shown in Fig.~\ref{dataset2confusion}. It can be seen that, while NeuroPath2Path distinguishes Motor-1, Visual-1, and Visual-2 quite well, the class of Somato-1 is significantly misclassified with Motor-1, Motor-2, and Visual-1, leading to a decline in the classification score.  

\begin{figure}[h]
\vspace{-.1cm}	
	\centering
	\includegraphics[width=6.6cm, height=3cm]{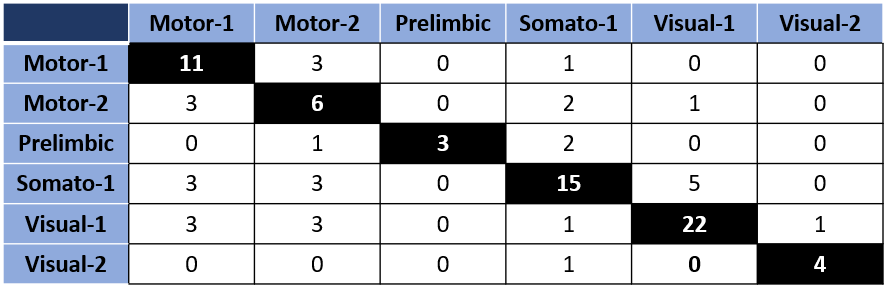}
	\caption{Confusion matrix of an instance of classification using dataset-1. The overall accuracy is $66\%$. Here, we set $L = 50$ and $K = 3$.}
	\label{dataset1confusion}
	\vspace{-.1cm}
\end{figure}

\begin{figure}[h]
\vspace{-.4cm}	
	\centering
	\includegraphics[width=9.6cm, height=4cm]{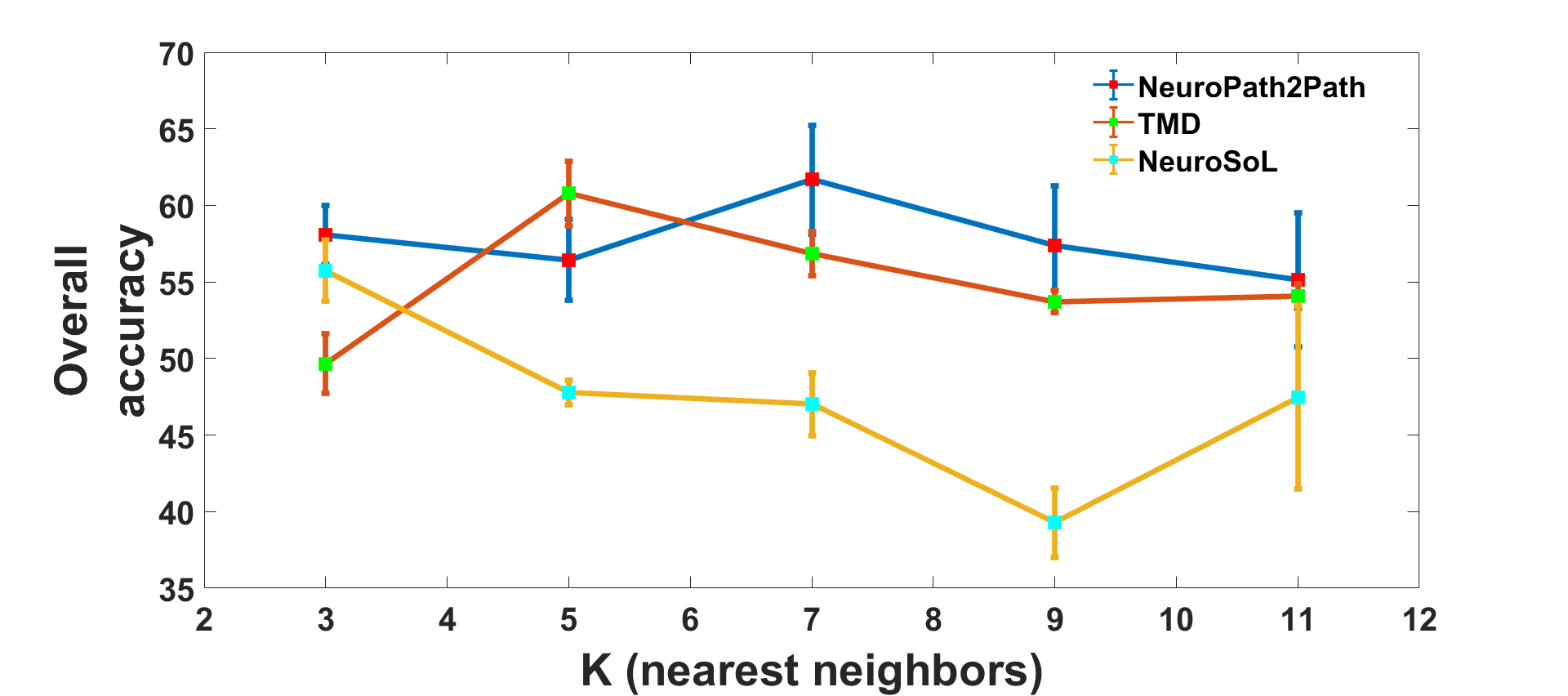}
	\caption{The figure shows comparative performance of NeuroPath2Path against TMD and NeuroSoL using different values of $K$ in K-NN classifier. At each $K$, we perform $5$ experiments for each of these methods and the associated scores are shown with the mean (colored square) and associated range of values.
	}
	\label{dataset1comp}
	\vspace{-.4cm}
\end{figure}

Next, in Fig.~\ref{dataset1comp}, we illustrate the comparative performance of NeuroPath2Path against TMD and NeuroSoL. The train to test ratio is set at $8:2$. NeuroSoL shows an erratic behavior as $K$ increases. TMD offers a consistent margin of classification accuracy per $K$. Here, at a given value of $K$, margin implies the difference between the maximum and minimum scores of $5$ experiments which are independently instantiated by randomly partitioning the dataset with $8:2$ train$:$test ratio. Fig.~\ref{dataset1comp} suggests that NeuroPath2Path achieves peak performance when $K$ is set as $7$, but with a noticeable margin. 
\begin{figure}[h]
\vspace{.1cm}	
	\centering
	\includegraphics[width=8.7cm, height=1.5cm]{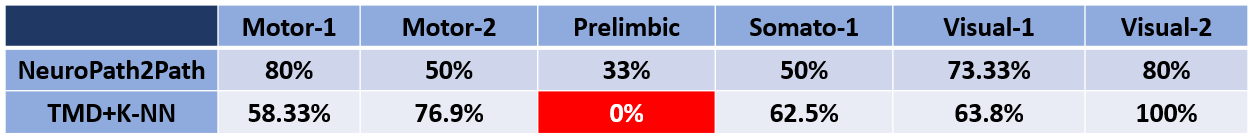}
	\caption{This figure shows one typical instance of classwise retrieval accuracy of NeuroPath2Path and TMD. NeuroPath2Path maintains almost consistent classwise performance. }
	\label{classwiseACC1}
	\vspace{-.4cm}
\end{figure}
To scrutinize the performance of TMD and NeuroPath2Path, we routinely inspect the class-wise retrieval accuracy, a crucial metric which is obscured in Fig.~\ref{dataset1comp} due to the averaging effect. The result is shown in Fig.~\ref{classwiseACC1}.
In a majority of cases, despite comparable overall classification scores, TMD tends to be affected by class imbalance, leading to significantly poor accuracy for few classes.

\subsection{Dataset-2 (Interclass)}
\begin{figure}[h]
\vspace{-.1cm}	
	\centering
	\renewcommand{\tabcolsep}{0.01cm}
\begin{tabular}{c}
	{\includegraphics[width=8.5cm, height=3.6cm]{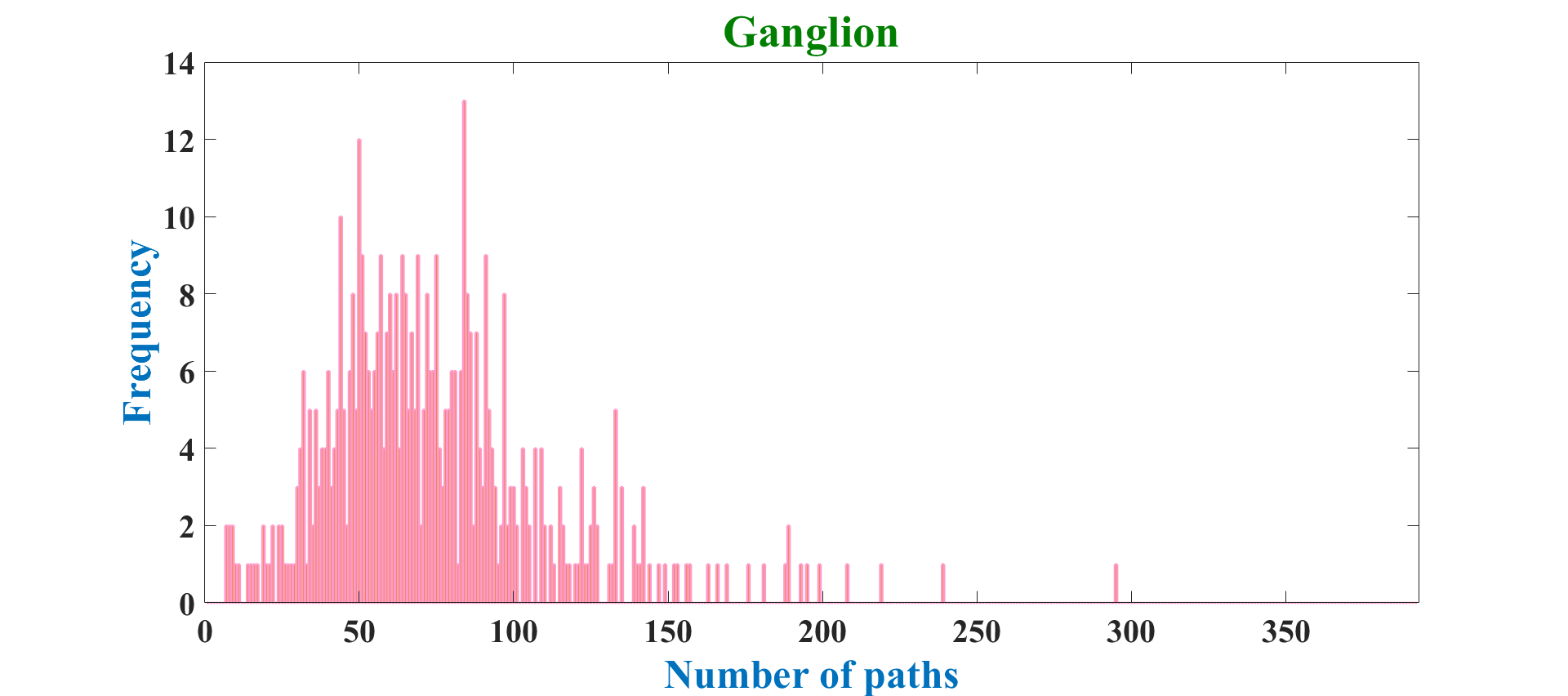}}\\
	{\includegraphics[width=8.5cm,height=3.6cm]{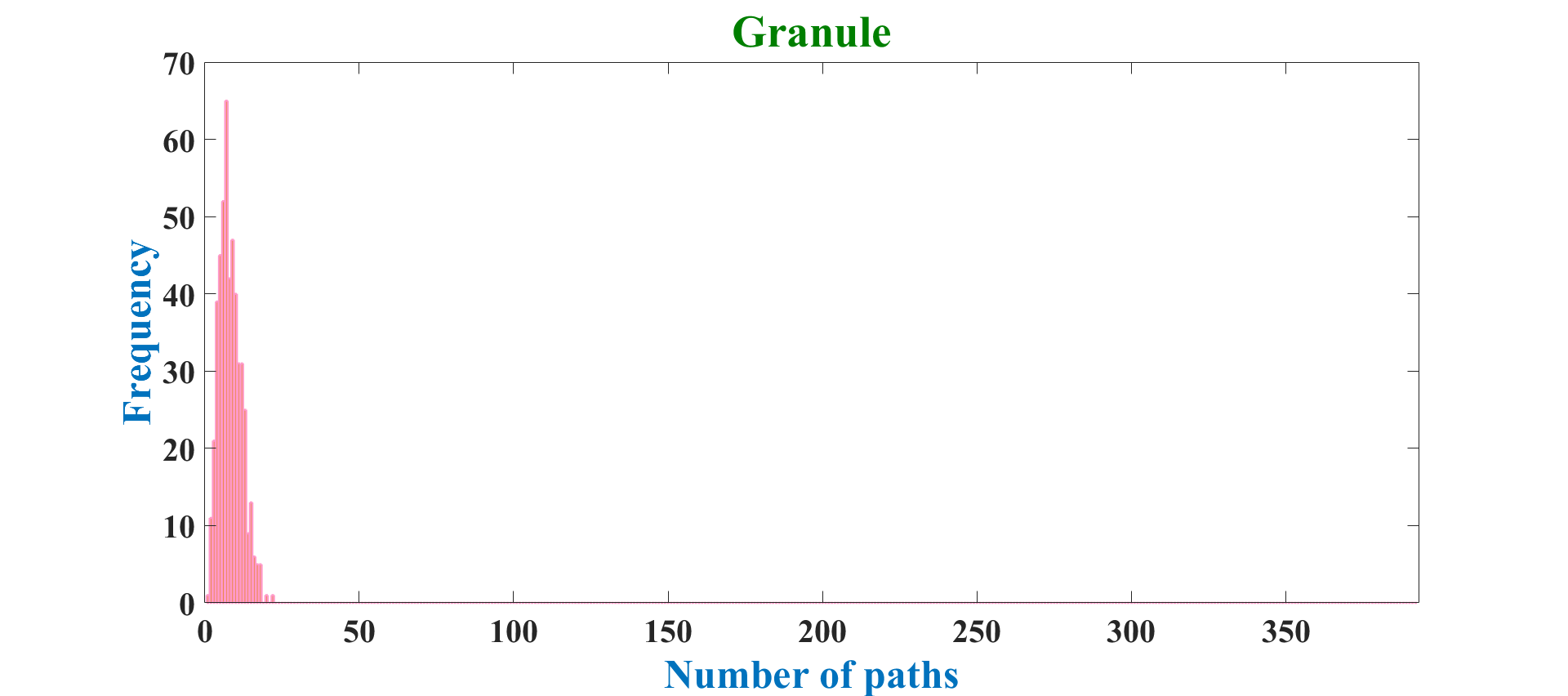}}\\
    {\includegraphics[width=8.5cm,height=3.6cm]{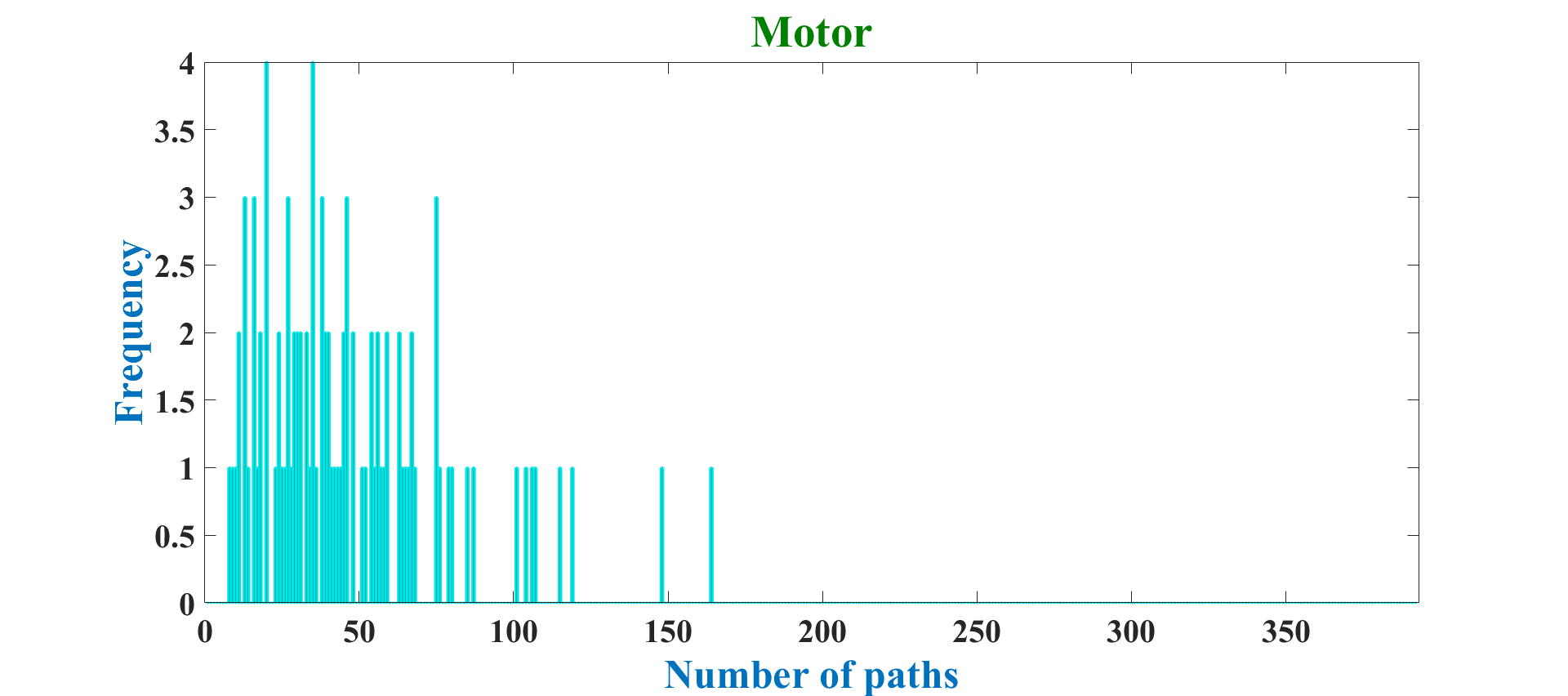}}\\
    {\includegraphics[width=8.5cm,height=3.6cm]{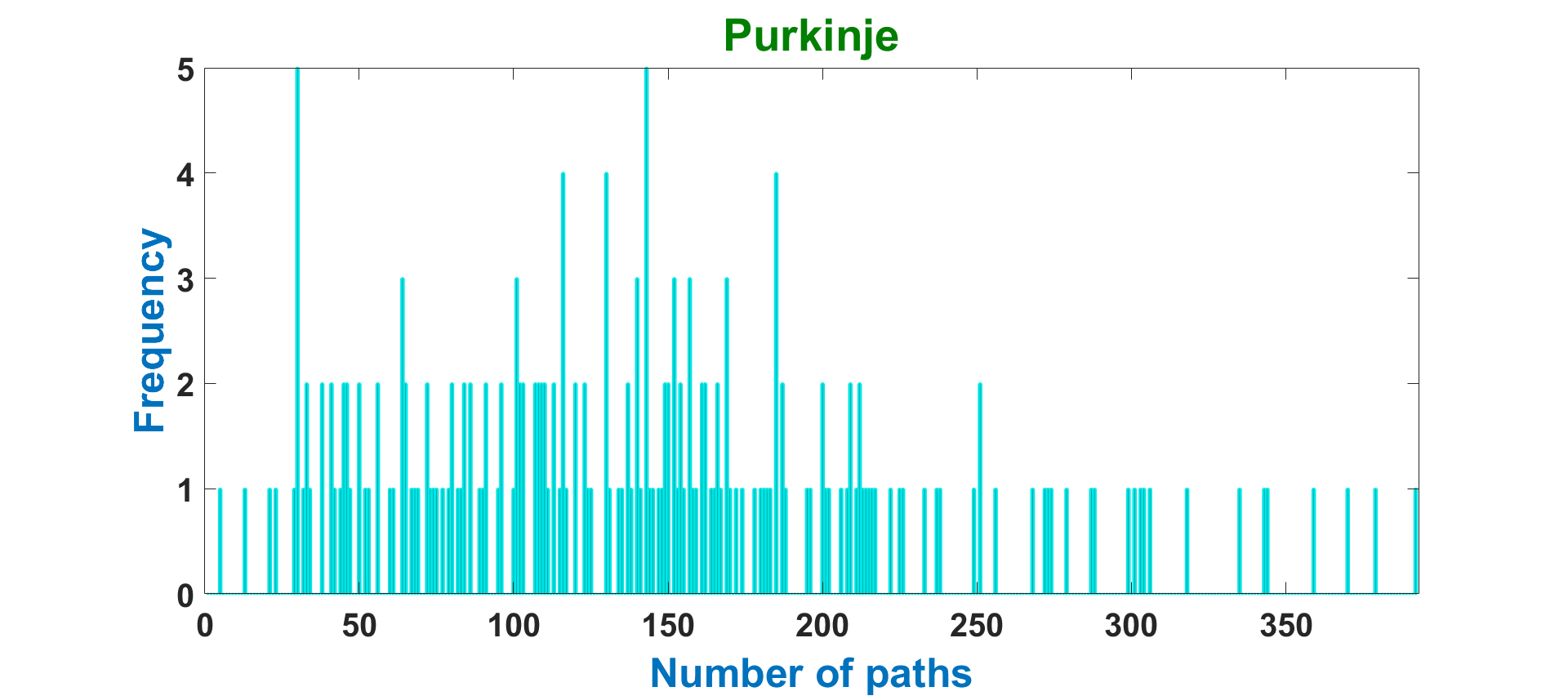}}\\
    {\includegraphics[width=8.5cm,height=3.6cm]{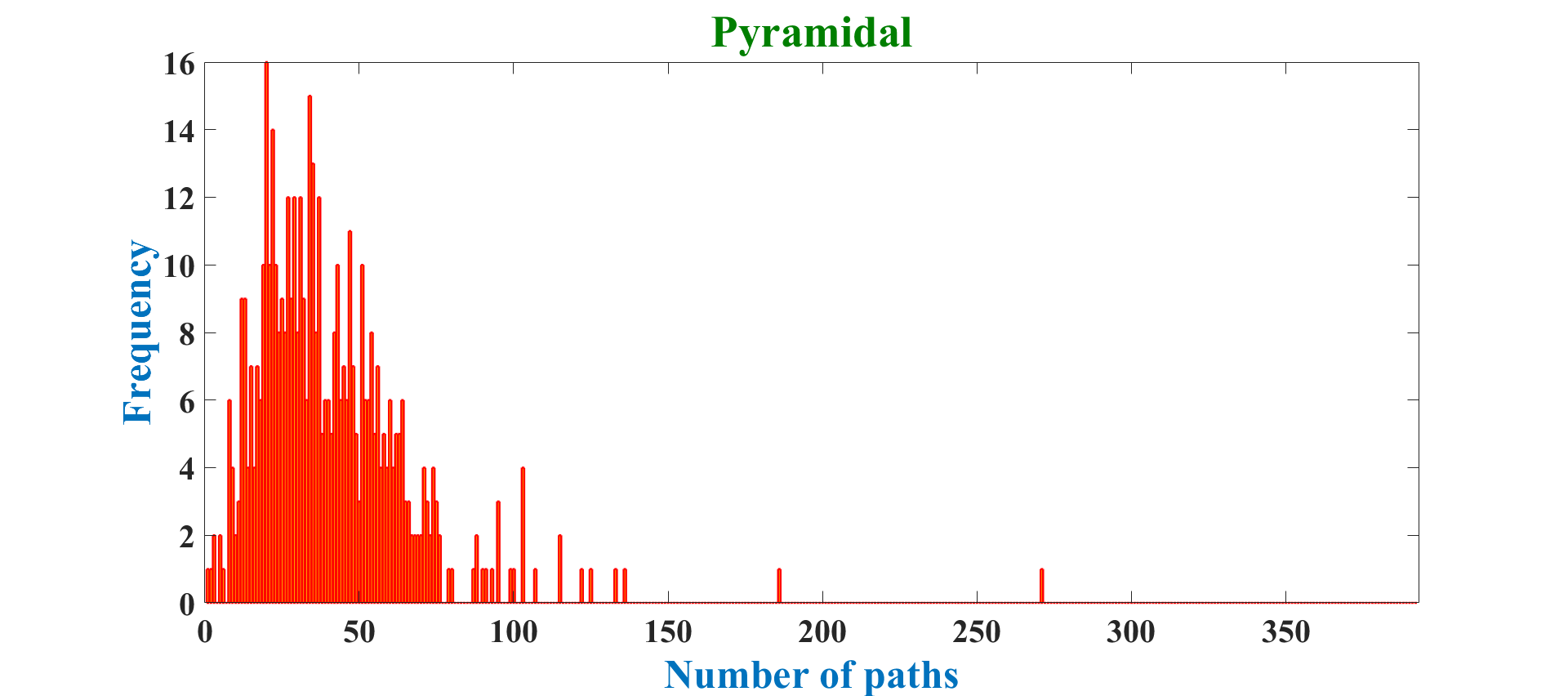}}\\
\end{tabular}
\caption{The figure shows the cell-specific distribution of the number of paths. It is observed that the distribution of paths in the case of Purkinje cells is approximately uniform. The remainder of the distributions are left-skewed. }
	\label{fig: pathdistCell}
	\vspace{-.8cm}
\end{figure}

The second dataset consists of 3D-reconstructed neurons that are traced from five major cell types of the mouse: ganglion, granule, motor, Purkinje, and pyramidal. We experiment with an imbalanced pool of $500$ ganglion cells, $490$ granule cells, $95$ motor cells, $208$ purkinje cells, and $499$ pyramidal cells, where the corresponding SWC files are obtained from the neuromorpho repository. The cell-specific distribution of paths is shown in Fig~\ref{fig: pathdistCell}.

For classification, we compute the important weights $\delta$ of each features, and due to space constraints, the $\delta$ values are enumerated in Table~\ref{importSet2} for pairwise classes and the case with all the classes taken together. The importance-weighted distance value, $\mu^{fg}$ in (\ref{dfg}) is used to compute the distance of a pair of neurons. We empirically find that the nonlinear transformation of $\mu^{fg}$, given by $\frac{1}{1+exp(-\mu^{fg})}$, yields an improved classification performance.  

\begin{table*}[ht]
\vspace{-.1cm}	
	\caption{Importance weight $\delta$ values for dataset-2}
	\label{importSet2}
	\centering
	\begin{tabular}{|l!{\vrule width 1.5pt}c|c|c|c|c|c|}
		\hline
		 & Tortuo$(\kappa)$ & Bifur-angle$(b)$ & Part-aym$(\alpha)$ &Concur$(C)$ & Seg-len$(\beta)$ & Diverg$(\lambda)$  \\\hline 
		Ganglion-Granule   &0.0278  &0.0138  &0.0241   &0.2102    &0.6727    &0.0533   \\\hline
		Ganglion-Motor &0.0464   &0.0797  &0.0918   &0.1865    &0.4730    &0.1226   \\\hline
		Ganglion-Purkinje   &0.27  &0.032  &0.1437   &0.1905    &0.2270    &0.1368   \\\hline
		Ganglion-Pyramidal  &0.0433  &0.0001   &0.0553   &0.5121    &0.3156    &0.0809   \\\hline
		Granule-Motor  &0.0356  &0.0491  &0.0372   &0.1449    &0.6697    &0.0636   \\\hline
		Granule-Purkinje &0.0147 &0.1218  &0.1875   &0.1073    &0.5285    &0.0402   \\\hline
		Granule-Pyramidal   &0.0453  &0.0305  &0.0372   &0.2231    &0.5390   &0.1248   \\\hline
		Motor-Purkinje  &0.0094  &0.4046  &0.0556   &0.0007    &0.5167    &0.0130   \\\hline 
		Motor-Pyramidal  &0.0223  &0.0737  &0.0465   &0.1275    &0.6650    &0.0650   \\\hline 
		Purkinje-Pyramidal &0.0219 &0.2201  &0.1088   &0.1528    &0.4223    &0.0740   \\\hline
		Overall         &0.1304  &0.0372  &0.0417   &0.1786    &0.4489    &0.1633   \\\hline 
	\end{tabular}
	\vspace{-.1cm}
\end{table*} 

\begin{figure}[t]
\vspace{-.1cm}	
	\centering
	\includegraphics[width=6.6cm, height=3cm]{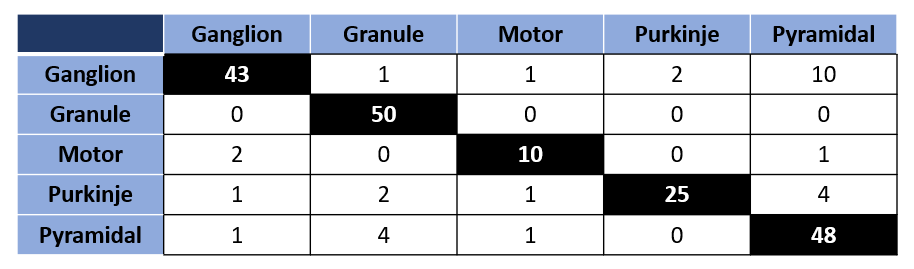}
	\caption{Confusion matrix of an instance of classification using Dataset-2. The overall accuracy is $85.02\%$. Here, we set $L = 50$ and $K = 9$. It can be seen from the matrix that one-fifth of ganglion cells are misclassified as pyramidal, leading to a decline in accuracy. However, granule cells are perfectly classified. }
	\label{dataset2confusion}
	\vspace{-.1cm}
\end{figure}

\begin{figure}[h]
\vspace{-.4cm}	
	\centering
	\includegraphics[width=9.6cm, height=5cm]{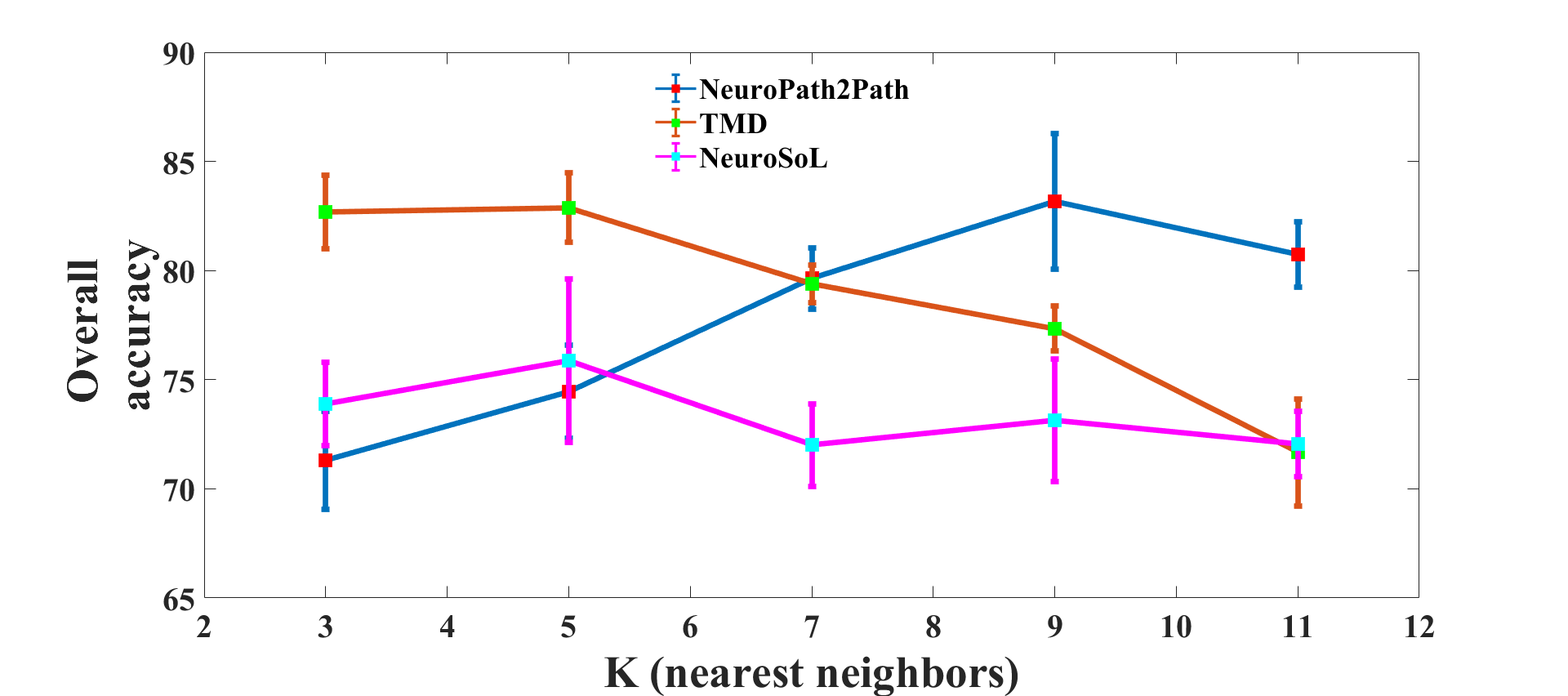}
	\caption{The figure shows comparative performance of NeuroPath2Path against TMD and NeuroSoL using different values of $K$ in K-NN classifier. At each $K$, we perform $5$ experiments for each of these methods and the associated scores are shown with the mean (colored square) and the range values.
	The profiles of TMD and NeuroPath2Path surprisingly appear to have opposite trends over $K$. NeuroPathPath hits the top accuracy of $86.2\%$ when $K=9$.}
	\label{dataset2comp}
	\vspace{-.1cm}
\end{figure}

With a train$:$test ratio as $8:2$, one instance of the confusion matrix, obtained by NeuroPath2Path is provided in Fig.\ref{dataset2confusion}.
We demonstrate the effectiveness of NeuroPath2Path over two state-of-the-art approaches - Topological Morphological Descriptor (TMD)~\cite{kanari2018topological} and NeuroSoL~\cite{batabyal2017neurosol}. For each value of $K$, we randomly partition the dataset $5$ times maintaining a constant $9:1$ ratio between the train and test datasets. In short, for every $K$, we obtain $5$ accuracy scores, which are plotted in Fig.~\ref{dataset2comp}.

\begin{figure}[h]
\vspace{-.1cm}	
	\centering
	\includegraphics[width=8.7cm, height=2.2cm]{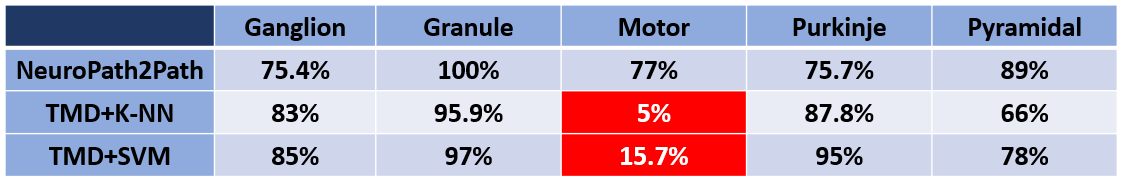}
	\caption{This figure shown the classwise retrieval accuracy of different methods including NeuroPath2Path. It is observed that by using TMD the retrieval accuracy of Motor cells shows minimal improvement when SVM is used. The imbalance in class adversely affects the classification accuracy. NeuroPath2Path maintains consistent class performance. }
	\label{classwiseACC}
	\vspace{-.1cm}
\end{figure}

TMD appears to be very consistent in accuracy and range scores, achieving an accuracy of $85.02\%$ when $K=5$. However, while computing the confusion matrices of the classification scores obtained by TMD, we notice that in the majority of instances, the correct classification of motor cells is abnormally low and approaches $0\%$ in some cases. It is important to notice that Dataset-2 has an imbalance in terms of the number of examples in each cell category, with motor cells containing the lowest ($95$) and ganglion cells containing the highest ($500$) number of examples. This fact is unobserved in Fig.~\ref{dataset2comp} due to the averaging effect. We adopt the metric, class-wise accuracy of retrieval, and present the results in Fig.~\ref{classwiseACC}. It is evident that NeuroPath2Path exhibits strong resilience against the class imbalance problem.      

\begin{figure}[h]
\vspace{-.1cm}	
	\centering
	\includegraphics[width=9.6cm,height=4.1cm]{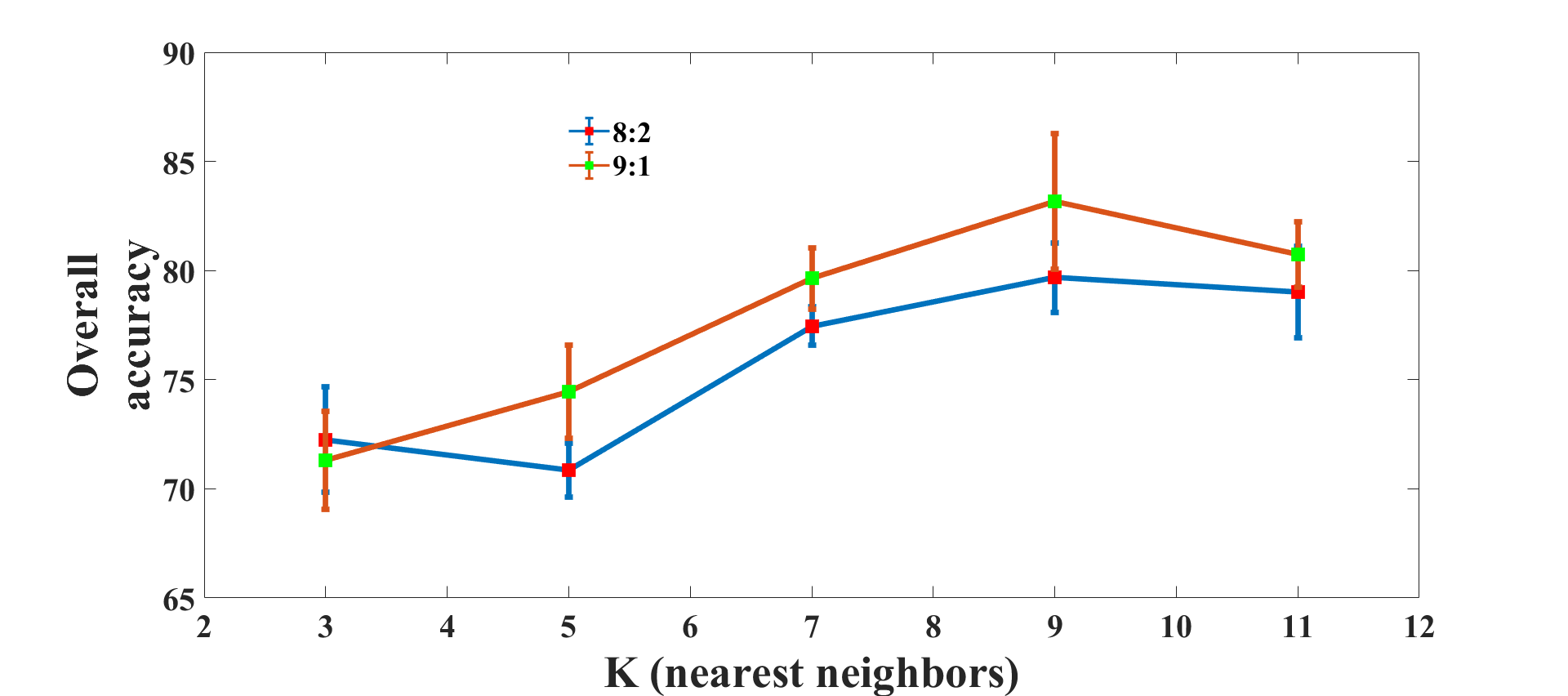}
	\caption{The performance of NeuroPath2Path on two different partitions, which are $9:1$ and $8:2$, of Dataset-2 is shown. $K=9$ is found to be a suitable candidate of K-NN classifier. }
	\label{data2ratio}
	\vspace{-.1cm}
\end{figure}

Similar to the train and test ratio of Dataset-1, we conduct experiments using the ratio of $0.1$ and $0.2$ separately. The classification scores are given in Fig.~\ref{data2ratio}.  

\section{Conclusion}
NeuroPath2Path follows a graph-theoretic approach that utilizes path-based modeling of neuron anatomy and provides a visualization tool by way of a geometric model that aids in performing continuous deformation between two neurons. NeuroPath2Path offers several advantages. The decomposition of a neuron into paths can be viewed as an assembly of individual circuits from the terminals to the soma, integrating semi-local features that act as path descriptors. Next, instead of subgraph matching, NeuroPath2Path does not leave a single path unassigned, culminating in a full-graph matching algorithm. The matching algorithm presents the notion of relative fractality and path correspondence, and incorporates physiological factors, such as decaying importance of features along the path and exploratory /competitive behavior for resource exploitation. 

NeuroPath2Path also precisely investigates the feasibility of algorithmic constraints (such as on Munkres algorithm) on the structural repertoire of neuronal arbors, and thereby enforcing criteria, such as hierarchy mismatch. During classification, NeuroPath2Path delivers resilience to the class imbalance problem. In the future, in order to explore the full potential of the approach beyond classification, we aim to augment NeuroPath2Path in two major domains - morphological analysis and structural transformation of microglia cells, and in progressive degradation of neuronal paths in neurodegenerative diseases.

\section*{Appendix}
\subsection*{Description of features}
We extract a set of discriminating features on each path $f_i\in\Gamma$ of $H$, which are bifurcation angle ($b_i$), concurrence ($C_i$), hierarchy ($\xi_i$), divergence ($\lambda_i$), segment length ($\beta_i$), tortuosity ($\kappa_i$), and partition asymmetry ($\alpha_i$).
\noindent\textbullet\hspace*{.2mm} \textit{Bifurcation angle} is a key morphometric that dictates the span and the spatial volume of an arbor. It is hypothesized that the span of an arbor at each level of bifurcation depends on the bifurcation of its previous level~\cite{lopez2011models, batabyal2018neurobfd, bielza2014branching}, suggesting the influence of Bayesian philosophy. This organizational principle is utilized in several stochastic generative models~\cite{lopez2011models} for the synthesis of specific neuron cell types. The sequence of bifurcation angles at bifurcation vertices located on a path of a neuron captures local geometry. For example, a sequence of non-increasing bifurcation angles from the root to the dendritic terminal of a path indicates the pyramidal shape geometry of the neuron. For a location with multifurcation, we use the maximum of the bifurcation angles computed using pairwise branches originated from that location towards the dendritic terminals.   

\noindent\textbullet\hspace*{.2mm} \textit{Concurrence, hierarchy and divergence} encode the effect of phenomenological factors, which are exploration (ex. Purkinje fanning out rostrocaudally) and competition (ex. retinal ganglion cells), that contribute in the growth of a neuron. The definition of concurrence and hierarchy are already given in section~\ref{path2path}. The divergence of a location on a path, $f_i$ is proportional to the repulsive force that the location experiences from its neighborhood path segments. Let $C_{f_i}$ be the sequence of concurrence values of the path $f_i \in \Gamma$ when one visits the locations from the root to the dendritic terminal. As an open curve, each path can be parameterized with the parameter $t\in[0,1]$. $C_{f_i}(t_s) = k;~t_s\in[0,1]$ indicates that $k (\le|\Gamma|)$ paths share the location $t_s$ on $f_i$. The divergence $\lambda$ of a location $f_i(t_s)$ is defined as $\lambda (f_i(t_s)) = $$ 1_{\{f_j~|~|f_j(t)-f_i(t_s)|\le \epsilon, f_j\neq f_i , f_j\nsucc f_i \}}$. Here, $1$ is the indicator function computing the number of such $f_j$s which follow the conditions $|f_j(t)-f_i(t_s)|\le\delta, f_j\neq f_i $ and $ f_j\nsucc f_i $. The first condition implies that a location of $f_j$ has to be in the $\epsilon$ neighborhood of $f_i$. $f_j\nsucc f_i$ indicates that the location of bifurcation at which $f_j$ deviates from $f_i$ does not appear after $f_i(t_s)$ on the path $f_i$.    
      
\noindent\textbullet\hspace*{.2mm} \textit{Tortuosity and partition asymmetry} are two important anatomical features of a neuron. Tortuosity refers to the amount of `zig-zag' or bending of a path. Let us take a segment on a path $f_i$ as $f_i([t_1,t_2]);~0\le t_1 < t_2 \le 1$. Let there be $m-1$ locations in $[t_1, t_2]$. The tortuosity of the segment is defined as $\kappa = \frac{\sum^{m}_{j=1}||f_i(t_{j+1}) - f_i(t_{j})||_2}{||f_i(t_2) - f_i(t_1)||_2}$ with $t_{m+1}=t_2$. Partition asymmetry accounts for how the size of a neuron tree varies within the neuron. We use a variant of caulescence, proposed in~\cite{brown2008quantifying}, as a measure of tree asymmetry. Caulescence at a bifurcation location is evaluated by way of $\alpha = \frac{|l-r|}{l+r}$, where $l$ is the size of the left tree and $r$ of the right tree of the bifurcation vertex. We define the size of a tree by the number of paths or equivalently the number of dendritic terminals. Note that the quantity $(l+r)+1$ is the concurrence value of the bifurcation vertex.

\begin{figure*}[h]
\vspace{.1cm}	
	\centering
	\renewcommand{\tabcolsep}{0.05cm}
\begin{tabular}{ccccc}
	{\includegraphics[width=3.0cm, height=2.5cm]{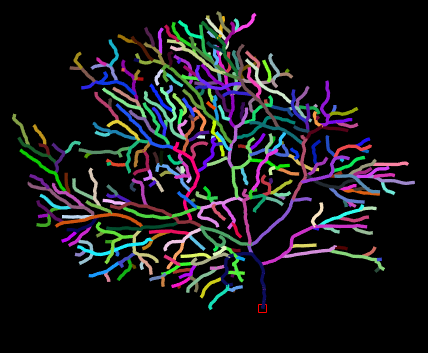}}&
	{\includegraphics[width=3.0cm,height=2.5cm]{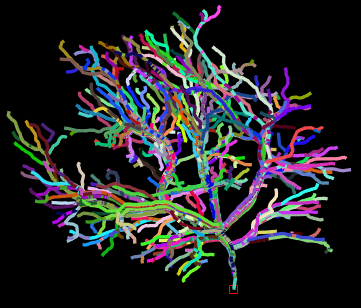}}&
    {\includegraphics[width=3.0cm,height=2.5cm]{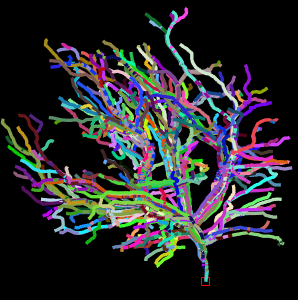}}&
    {\includegraphics[width=3.0cm,height=2.5cm]{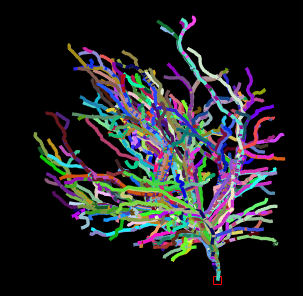}}&
    {\includegraphics[width=3.0cm,height=2.5cm]{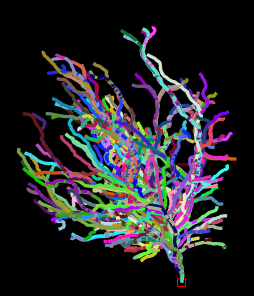}}\\
    {\includegraphics[width=3.0cm,height=2.5cm]{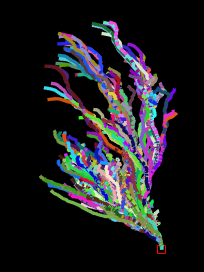}}&
    {\includegraphics[width=3.0cm,height=2.5cm]{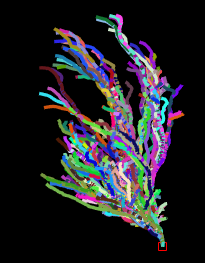}}&
    {\includegraphics[width=3.0cm,height=2.5cm]{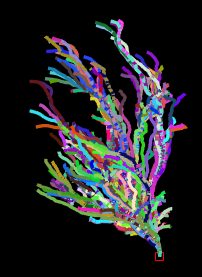}}&
    {\includegraphics[width=3.0cm,height=2.5cm]{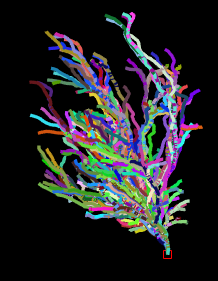}}&
    {\includegraphics[width=3.0cm,height=2.5cm]{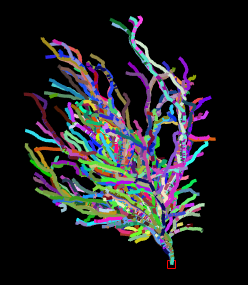}}\\
    {\includegraphics[width=3.0cm,height=2.5cm]{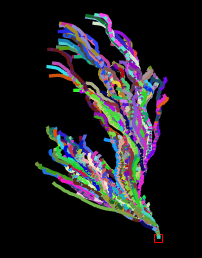}}&
    {\includegraphics[width=3.0cm,height=2.5cm]{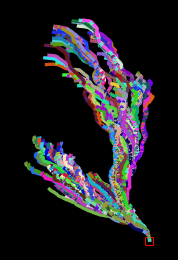}}&
    {\includegraphics[width=3.0cm,height=2.5cm]{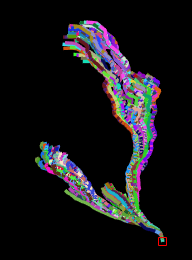}}&
    {\includegraphics[width=3.0cm,height=2.5cm]{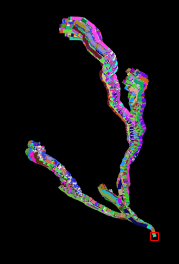}}&
    {\includegraphics[width=3.0cm,height=2.5cm]{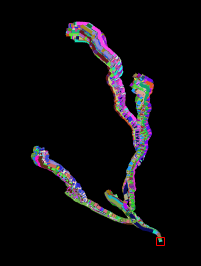}}\\
    {\includegraphics[width=3.0cm,height=2.5cm]{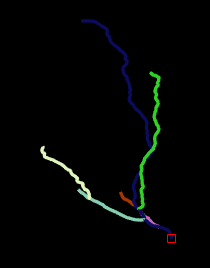}}&
    {\includegraphics[width=3.0cm,height=2.5cm]{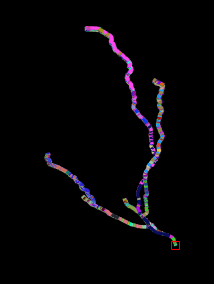}}&
    {\includegraphics[width=3.0cm,height=2.5cm]{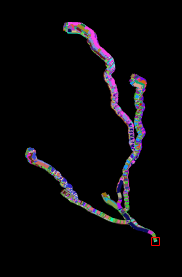}}&
    {\includegraphics[width=3.0cm,height=2.5cm]{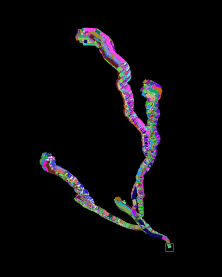}}&
    {\includegraphics[width=3.0cm,height=2.5cm]{Pur2Gr15.PNG}}\\
\end{tabular}
\caption{This gallery of images captures the progressive evolution of paths from a Purkinje neuron to a granule one. The granule neuron ($426.5\mu m^3$ volume) is procured from the hippocampus (dentate gyrus) of a $5$ months old mouse, containing $6$ rooted paths. The Purkinje cell ($13094\mu m^3$ volume) is sampled from the cerebellar cortex of a $35$ day-old mouse, containing $304$ paths. The evolution is represented in multiple arrays such that the ODD rows are read in the left-to-right and the EVEN rows are read in the right-to-left fashion. In the first column, the top image is the granule cell and the bottom shows the pyramidal cell. Volume-wise, the Purkinje neuron is significantly larger than the granule neuron. However, they are scaled for visualization.}
	\label{fig: purgrmorph}
	\vspace{-.1cm}
\end{figure*}

\subsection*{Weight determination}
Let the combined distance vector containing the individual feature distances be $D^{fg} = [d(b^{fg})~d(C^{fg})~d(\lambda^{fg})~$ $d(\kappa^{fg})~d(\beta^{fg})~d(\alpha^{fg})]^T$.
The corresponding unknown weight vector is $\delta = [\delta_1,...\delta_6]$.
While comparing two neurons of sizes $N$ and $M$ with $N<=M$, the distance computation after applying the Munkres algorithm repeatedly will produce $M$ pairs of paths, indicating $M$ such $D^{fg}$s. The desired characteristic of each component of $\delta$ is positivity. In addition, we enforce $\sum\delta_i~=~1$, implying a probability estimate. $\delta_i$ thus indicates the relative importance of the feature $\upsilon_i$. 


We adopt the constrained \textit{maximizing-interclass} \textit{-minimizing-intraclass} distances strategy to find our desired $\delta$. Mathematically,
\begin{eqnarray}
\label{OptimEq}
\delta^{opt} &=& -\argminC_{\delta}\frac{1}{2}\delta^T\Big(\sum^S_{\substack{i,j=1\\
               i<j}}\sum_{k=1}^{N_i}\sum_{l=1}^{N_j}\sum_{z=1}^{M_{kl}}D^{z}(D^z)^T\Big)\delta \nonumber\\
               && + \delta^T\Big(\sum_{i=1}^S\tau_i\sum^{N_i}_{\substack{k,l=1\\ k\neq l}}\sum_{z}^{M_{kl}}D^z(D^z)^T\Big)\delta \nonumber\\
               && - \omega_1log\delta + \omega_2\big(\sum_{i=1}^6\delta_i - 1\big)
\end{eqnarray}
The first term in the above equation encompasses all the distances between neurons from pairwise classes. The second term encodes the intraclass distances, implying the distances between neurons for each class. The third term enforces positivity of each weight $\delta_i$. This is a logarithmic barrier penalty term that restricts the evolution of $\delta$ at intermediate iterations to the region where $\delta > \bar{0}$. The last term accounts for the probabilistic interpretation of $\delta$. $S$ is the number of classes. 

Eq.~\ref{OptimEq} is solved by using gradient descent. The equation and its derivative can be simply written as,
\begin{eqnarray}
L(\delta) &=& -\frac{\delta^T\Pi\delta}{2} + \delta^T\sum_{i=1}^S\frac{\tau_i\Pi_i}{2}\delta-\omega_1log\delta+\omega_2\big(\delta\boldsymbol{1}-1\big)\nonumber\\
\frac{dL}{d\delta} &=& -\Pi\delta +\sum_{i=1}^S\tau_i\Pi_i\delta-\frac{\omega_1}{\delta}+\omega_2
\end{eqnarray}
We use this derivative term in the following algorithm~\ref{alg1} to obtain optimal $\delta$. 
\begin{algorithm}
\caption{Find $\delta$}
\label{alg1}
\KwData{$\Pi,\Pi_1, \Pi_2, ...,\Pi_S$ (for all classes)\;}
Initialization: $\delta_{cur}, \delta_{tmp},\tau_1,\tau_2,...,\tau_S, \omega_1,\omega_2$, Iter, tol, $\eta$\;
\While{iter $<$ Iter}{
    \While{$||\delta_{cur}-\delta_{tmp}||_2 < tol$}{
        $D\longleftarrow\sum_{k = 1}^S\tau_k\Pi_k$\;
        $Der\longleftarrow-\Pi\delta+D\delta-\frac{\omega_1}{\delta}+\omega_2$\;
        $\delta_{curr}\longleftarrow\delta_{tmp}$\;
        $\delta_{curr}\longleftarrow\delta_{curr}/(\delta_{curr}\boldsymbol{1}$)\;
        $\delta_{tmp}\longleftarrow\delta_{tmp}-\eta Der$\;
    }

    $iter\longleftarrow iter+1$\;
    $\omega_1\longleftarrow\omega_1/2$\;
    $\omega_2\longleftarrow2\omega_2$\;
    $\tau_i\longleftarrow1.1\tau_i\forall i\text{(more intraclass compaction)}$\;

}
\end{algorithm}

\subsection*{Distance between neurons}
The algorithm to find distance between a pair of neurons consists of four stages - finding self-similarity (routine-1), remaining path assignment (routine-2), finding pairs with hierarchy mismatch (routine-3) and reassignment of the defective pairs (routine-4).   

\begin{algorithm}
\caption{Find $\chi$}
\label{alg2}
\KwData{$\mathcal{D}, n_1,n_2~(n_1\leq n_2)$, Inf\;}
\KwOut{pairList, dstSum}
Initialization: $dstSum\longleftarrow 0$, $pairTmp\longleftarrow [1:n_2]^T$, $count\longleftarrow1$, $pairList\longleftarrow [~]$, $\mathcal{D}_c\longleftarrow\mathcal{D}$, $dstB\longleftarrow[~]$\;
\textbf{Routine 1:} (Self-similarity)\\
\While{$count~\leq~\lfloor\frac{n2}{n1}\rfloor$}{
    $\hat{\mathcal{D}}\longleftarrow \mathcal{D}$ (append rows with Inf)\;
    $pairL,dst \longleftarrow Munkres (\hat{\mathcal{D}})$\;
    $dstSum\longleftarrow dstSum+dst$\;
    $dstB\longleftarrow dstB\cup dst$\;
    $pairTmp\longleftarrow pairTmp\setminus pairL[1:n_1]$ ($(n_2-n_1)~\text{dummy}$)\;
    $pairList\longleftarrow pairList \cup pairTmp[pairL[1:n_1]]$\;
    $\mathcal{D}\longleftarrow \mathcal{D}[1:n_1,pairTmp]$\;
    $count\longleftarrow count+1$\;
}
\textbf{Routine 2:} (Remaining $n_2 - \lfloor\frac{n2}{n1}\rfloor n_1$ paths)
$T\longleftarrow n_2 - \lfloor\frac{n2}{n1}\rfloor n_1$\;
$\mathcal{D}\longleftarrow\mathcal{D}^T$\;
$\hat{\mathcal{D}}\longleftarrow \mathcal{D}$ (append rows with Inf)\;
$pairL,dst \longleftarrow Munkres (\hat{\mathcal{D}})$\;
$dstSum\longleftarrow dstSum+dst$\;
$dstB\longleftarrow dstB\cup dst$\;
$pairList\longleftarrow pairList \cup pairTmp[pairL[1:T]]$\;
\vspace*{0.1cm}
\textbf{Routine 3:} (Hierarchy mismatch)\\
$dg\longleftarrow Hist[pairList[:,1]]$
$deg\longleftarrow Hist[pairList[:,2]]$ (second column)\;
$md\longleftarrow Median[dstB],~sd\longleftarrow SD[dstB]$\;
$sk\longleftarrow Skewness[dstB]$\;
\uIf{$sk > 0$}{
    $IX\longleftarrow where[dstB > md+sd]$
}
$misalign2\longleftarrow[~]$\;
$listU \longleftarrow [~]$\;
\For{$k\gets 1$ \KwTo $length[IX]$}{
    $pvt\longleftarrow pairList[k,1:2]$\;
    $h1\longleftarrow H[pvt[1]]$,~$h2\longleftarrow H[pvt[2]]$(max hierarchy)\;
    \uIf{$|h1-h2| > \frac{max[h1,h2]}{2}$}{
        \uIf{$deg[pvt[2]]<2$}{
            $misalign2\longleftarrow misalign2\cup pvt[2]$\;
            $misalign1\longleftarrow misalign1\cup pvt[1]$\;
            $dstSum\longleftarrow dstSum-dstB[k]$\;
            $listU\longleftarrow listU\cap pairList[k,:]$\;
        }
    }
}
\uIf{$listU \neq [~]$}{
    $Del~pairList[listU]$
}
\end{algorithm}
\begin{algorithm}
\textbf{Routine 4:} (Reassignment)\\
\For{$k=1$ \KwTo $length[misalign2]$}{
    $col\longleftarrow misalign2[k]$\;
    $row\longleftarrow \argminC_{row}\mathcal{D}_c[:,col]$\;
    $dstSum\longleftarrow dstSum+\mathcal{D}_c[row,col]$\;
    $pairList\longleftarrow pairList\cup [row,col]$\;
}
\For{$k=1$ \KwTo $length[misalign1]$}{
    \uIf{$dg[misalign1[k]] > 1$}{
        $row\longleftarrow misalign1[k]$\;
        $col\longleftarrow \argminC_{col}\mathcal{D}_c[row,:]$\;
        $dstSum\longleftarrow dstSum+\mathcal{D}_c[row,col]$\;
        $pairList\longleftarrow pairList\cup [row,col]$\;
    }
}
\end{algorithm}

\bibliographystyle{spmpsci}
\bibliography{NeuroP2P}




\end{document}